\documentclass{article}
\usepackage[brazil]{babel}
\usepackage[utf8]{inputenc}
\usepackage[margin=.8in]{geometry}
\usepackage{graphicx}
\usepackage{amsmath,lettrine}
\usepackage{gastex}

\newtheorem{definicao}{Definição}
\newtheorem{lema}{Lema}      
\newtheorem{fato}{Fato}      

\newcommand{\tr}{\mbox{\rm\emph{trie}}}
\newcommand{\trc}{\mbox{\rm\emph{trie} condensada}}

\newcommand{\rk}[1]{\mbox{\rm{rank}}(#1)}
\newcommand{\sk}[1]{\mbox{\rm{sketch}}(#1)}

\newcommand{\trs}[1]{\mbox{\rm{BuscaTrie}}(#1)}
\newcommand{\dlt}[2]{\Delta(#1,#2)}

\begin{document}

\title{Ordenação Baseada em Árvores de Fusão}

\author{Rogério H. B. de Lima \\ \small Instituto de Ciência e Tecnologia \\ \small Unifesp \and Luis A. A. Meira \\ \small Faculdade de Tecnologia \\ \small Unicamp}

\maketitle


\begin{abstract}
 O problema da ordenação é sem dúvida um dos mais estudados na Ciência da Computação. No escopo da computação moderna, depois de mais de 60 anos de estudos, ainda existem muitas pesquisas que objetivam o desenvolvimento de algoritmos que solucionem uma ordenação mais rápida ou com menos recursos comparados a outros algoritmos já conhecidos. Há vários tipos de algoritmos de ordenação, alguns mais rápidos, outros mais econômicos em relação ao espaço e outros com algumas restrições com relação à entrada de dados. O objetivo  deste trabalho é  explicar a estrutura de dados Ávore de Fusão, responsável pelo primeiro algoritmo de ordenação com tempo inferior a $ n \lg n $, tempo esse que criou certa confusão, gerando uma errada crença de ser o menor possível para esse tipo de problema.\\
\end{abstract}

\section{Introdução}

O problema da ordenação é talvez o mais estudado da Ciência da Computação.
A sua utilização está implícita em etapas intermediárias de quase todos os programas existentes, como
 banco de dados, planilhas, multimídia etc. Além disso, a ordenação é estudada pela computação há mais de setenta anos.
 O algoritmo \textit{Merge Sort}, largamente utilizado nos dias de hoje, 
 foi proposto por Von Neumann em 1945~\cite{cormen}.
 
 O problema da ordenação  consiste em receber uma sequência de $n$ números como entrada $A= (a_1, \ldots , a_n)$.
 A solução consiste em uma  permutação não decrescente  $A'= (a'_1, \ldots ,a'_n)$.
Apesar deste trabalho focar em números inteiros, a extensão para racionais, ponto flutuante e cadeias de caracteres tende a ser direta.

Todos os algoritmos de ordenação apresentam características que os fazem mais ou menos competitivos com relação a outros.
 Algumas destas características são o tipo de ordenação, estável ou não estável, utilização de espaço extra para a execução
  do algoritmo, tempo de ordenação. Alguns algoritmos podem ser mais rápidos que outros dependendo do tipo de entrada de dados. Por exemplo, a ordenação por seleção tende a ser vantajosa quando $n$ é pequeno. A ordenação por inserção
  tende a ser rápida quando o vetor está parcialmente ordenado. A ordenação por contagem é vantajosa quando 
 a  diferença entre o maior e menor elemento  é limitada.

Os algoritmos de ordenação  mais conhecidos
são baseados em comparações, como o \textit{Merge Sort, Heap Sort, Insertion Sort} e \textit{Quick Sort} e os baseados em contagem, como por exemplo \textit{Counting Sort, Bucket Sort} e o \textit{Radix Sort}. Os algoritmos baseados em contagem necessitam de uma sequência de entrada com algumas restrições. 
Quando essas restrições são satisfeitas, esses algoritmos podem resolver o problema da ordenação em tempo linear.

Existe um limite de tempo inferior  de $ \Omega(n\lg n) $ comparações para algoritmos de ordenação~\cite{knuth}. Este limite é baseado em uma árvore de decisão com $n!$ folhas, cada uma representando uma permutação do vetor de entrada. Cada permutação é uma candidata a solução. Dado que uma comparação pode distinguir dois ramos de árvore, serão necessários, no mínimo, $\lg(n!) = \Theta(n\lg n) $ comparações, no pior caso, para ordenar um vetor através de um algoritmo de ordenação baseado em comparações.
Este limite inferior foi mal interpretado, gerando uma falsa crença em parte da comunidade de que ordenação é um problema $\Omega(n\lg n)$. 
Tal limite não se aplica, por exemplo, a algoritmos que usam outras operações além de comparações durante a ordenação. A 
ordenação por contagem é capaz de ordenar um vetor sem realizar nenhuma comparação entre elementos de $S$.

O algoritmo analisado neste trabalho é baseado em comparações e faz $\Theta(n \lg n)$ comparações, porém são comparados
$(\lg n)^{1/5}$ números em $O(1)$. Ou seja, são efetuadas múltiplas operações em tempo constante.
O parágrafo a seguir foi extraído de~\cite{cormen}:

\begin{quote}
``O caso de ordenar \textit{n} inteiros de \textit{w} bits no tempo $o(n\lg n) $ foi considerado por muitos pesquisadores. Vários resultados positivos foram obtidos, cada um sob hipóteses um pouco diferentes a respeito do modelo de computação e das restrições impostas sobre o algoritmo. Todos os resultados pressupõem que a memória do computador está dividida em palavras endereçáveis de \textit{w} bits.~\cite{fredman} introduziram a estrutura de dados de Árvore de Fusão e a empregaram para ordenar \textit{n} inteiros no tempo $O(n\lg n/\lg\lg n) $. Esse limite foi aperfeiçoado mais tarde para o tempo $O(n\sqrt{\lg n}) $ por~\cite{arne1}. Esses algoritmos exigem o uso de multiplicação e de várias constantes pré-calculadas.~\cite{arne2} mostraram como ordenar \textit{n} inteiros no tempo $O(n\lg\lg n) $ sem usar multiplicação, mas seu método exige espaço de armazenamento que pode ser ilimitado em termos de \textit{n}. Usando-se o hash multiplicativo, é possível reduzir o espaço de armazenamento necessário para $O(n) $, mas o limite $O(n\lg\lg n) $ do pior caso sobre o tempo de execução se torna um limite de tempo esperado. Generalizando as árvores de pesquisa exponencial de~\cite{arne1},~\cite{thorup} forneceu um algoritmo de ordenação de tempo $O(n(\lg\lg n)^2) $ que não usa multiplicação ou aleatoriedade, e que utiliza espaço linear. Combinando essas técnicas com algumas idéias novas,~\cite{han} melhorou o limite para ordenação até o tempo $O(n\lg\lg n\lg\lg\lg n) $. Embora esses algoritmos sejam inovações teóricas importantes, todos eles são bastante complicados e neste momento parece improvável que venham a competir na prática com algoritmos de ordenação existentes".
\end{quote}

Resultados: O algoritmo de ordenação $O(n\lg n/\lg\lg n)$ analisado neste trabalho 
é conhecido na literatura. Nossa contribuição consiste em detalhar a estrutura de dados Árvore de Fusão e  o 
algoritmo de ordenação  $O(n\lg n/\lg\lg n)$ proposto por~\cite{fredman}.

\subsection{Modelo Computacional}

Considere um computador  que trabalhe com palavras de $w$ bits. Este computador é capaz 
ralizar perações elementares como soma, subtração, multiplicação, divisão e resto de inteiros com $w$ bits em tempo constante. 
 Um computador de 64 bits tem capacidade de processar 64 bits em tempo constante, por exemplo.

O caso geral da ordenação  trata de inteiros com precisão arbitrária. Para um inteiro com $mw$ bits  são necessários $m$ acessos à memória antes de ler o número por completo. 
Este trabalho trata do caso restrito da ordenação onde os   números são  inteiros com $w$ bits.
 Tais números estão no intervalo $\{-2^{w-1},\ldots,2^{w-1}\}$ armazenados como inteiros binários, com 1 bit para o sinal.  

Este trabalho considerou um modelo computacional capaz de ler e escrever qualquer posição de memória em tempo constante, conhecido como memória RAM. 
O modelo de uma memória RAM é aceitável, apesar da coexistir com o modelo de  memória de acesso sequencial. Neste último, 
é necessário movimentar a fita até a posição desejada antes da leitura, gastando tempo linear para ler um inteiro.
  O algoritmo \emph{Merge  Sort} é famoso 
por manter a complexidade $O(n\lg n)$ mesmo num modelo de memória sequencial.

É razoável assumir que o computador é capaz de processar  $w=\log n$ bits tem tempo constante.
Se temos $n$ inteiros de $w$ bits na memória, o maior endereço de memória terá pelo menos $\lg n$ bits. Assumir que é possível 
acessar qualquer posição de memória em tempo constante equivale ao computador processar  endereços de $\lg n$ bits
em tempo constante.
Observe que o número de bits do problema é $nw \geq n \lg n$. Isto significa que uma operação para cada bit é $\Omega(n\lg n)$.

Para melhor compreensão da ordenação, primeiro será mostrado como ordenar $n$ números utilizando a estrutura de uma árvore B. A Árvore de Fusão, proposta por~\cite{fredman}, é uma árvore B modificada.

\section{Árvores B}

\begin{figure}[htb]
	\begin{center}
	\begin{picture}(50,50)(0,0)
	\node[Nadjust=wh,linecolor=white](CR)(0,26){\footnotesize $\vdots$} 
	\node[Nadjust=wh,linecolor=white](CR)(22,26){\footnotesize $\vdots$} 
	\node[Nadjust=wh,linecolor=white](CR)(50,26){\footnotesize $\vdots$} 	
	\node[Nadjust=wh,Nmr=1](A)(25,45){\footnotesize$s_1~|\ldots|~s_{B-1}$} 
	\node[Nadjust=wh,Nmr=1](B1)(0,30){\footnotesize $s_1~|\ldots|~s_{B-1}$} 
	\node[Nadjust=wh,Nmr=1](B2)(22,30){\footnotesize $s_1~|\ldots|~s_{B-1}$} 
	\node[Nadjust=wh,Nmr=1](B3)(50,30){\footnotesize $s_1~|\ldots|~s_{B-1}$} 
	\node[Nadjust=wh,linecolor=white](BR)(36,30){\footnotesize $\dots$} 
        \node[Nadjust=wh,linecolor=white](CR)(36,5){\footnotesize $\dots$} 
	\drawedge[ATnb=0,AHnb=1](A,B1){} 
	\drawedge[ATnb=0,AHnb=1](A,B2){} 
	\drawedge[ATnb=0,AHnb=1](A,B3){} 
	\node[Nadjust=wh,Nmr=1](D)(25,20){\footnotesize$s_1~|\ldots|~s_{B-1}$} 
	\node[Nadjust=wh,Nmr=1](C1)(0,5){\footnotesize $s_1~|\ldots|~s_{B-1}$} 
	\node[Nadjust=wh,Nmr=1](C2)(22,5){\footnotesize $s_1~|\ldots|~s_{B-1}$} 
	\node[Nadjust=wh,Nmr=1](C3)(50,5){\footnotesize $s_1~|\ldots|~s_{B-1}$} 
	\drawedge[ATnb=0,AHnb=1](D,C1){} 
	\drawedge[ATnb=0,AHnb=1](D,C2){} 
	\drawedge[ATnb=0,AHnb=1](D,C3){} 	 
	\end{picture}	
	\caption{Estrutura de uma árvore B completa.}
	\label{fig_arvoreB}
	\end{center}
\end{figure}
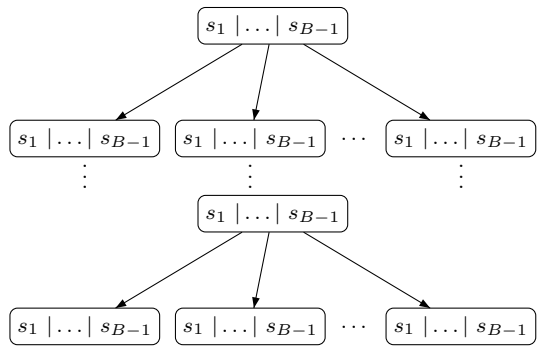

Árvores B são árvores de pesquisa balanceadas com grau $t$, onde $\frac{B}{2}~\leq~t~\leq~B$, para $B$ constante. Cada nó contém no mínimo $\frac{B}{2}$ e no máximo  $B$ filhos, com exceção das folhas, que não contém nenhum filho e do nó raiz, que não apresenta restrição de número mínimo de filhos. Cada nó da árvore possui no mínimo $\frac{B}{2}-1$ chaves e no máximo $B-1$ chaves ordenadas, e todas as folhas se encontram na mesma altura. Veja que uma nó com grau 4 possui 
 3 chaves. A Figura~\ref{fig_arvoreB} ilustra uma árvore B completa, onde todos os nós possuem $B - 1$ chaves.

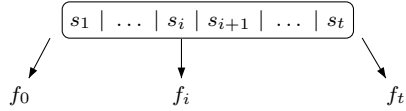
\begin{figure}[htb]
	\begin{center}
		\begin{picture}(50,15)(0,0)
	\node[Nadjust=wh,Nmr=1,linecolor=white](b1)(4.5,8.5){} 
		\node[Nadjust=wh,Nmr=1,linecolor=white](b2)(21.5,8.5){} 
			\node[Nadjust=wh,Nmr=1,linecolor=white](b3)(45,8.5){} 	
	\node[Nadjust=wh,Nmr=1](A)(25,10){\footnotesize$s_1~|~\ldots~|~s_i~|~s_{i+1}~|~\ldots~|~s_{t}$} 
	\node[Nadjust=wh,Nmr=1,linecolor=white](f1)(0,0){\footnotesize$f_0$} 
	\node[Nadjust=wh,Nmr=1,linecolor=white](f2)(21.5,0){\footnotesize$f_i$} 
	\node[Nadjust=wh,Nmr=1,linecolor=white](f3)(50,0){\footnotesize$f_{t}$} 
	\drawedge[ATnb=0,AHnb=1](b1,f1){} 
	\drawedge[ATnb=0,AHnb=1](b2,f2){} 
	\drawedge[ATnb=0,AHnb=1](b3,f3){} 	
	\end{picture}	\caption{Estrutura de um nó de uma árvore B}
	\label{fig_noArvoreB}
	\end{center}
\end{figure}

Além disso, a árvore B respeita a seguinte propriedade: cada nó não raiz  possui $t$ elementos elementos ordenados $S=(s_1, \ldots, s_t)$. 
Cada nó  não folha e não raiz possui  $t+1$ filhos
 ($f_0,\ldots, f_t)$ onde cada filho é uma árvore B.
 Os elementos na árvore $f_0$ são menores que $s_1$.
 Os elementos em $f_i$ são maiores que $s_i$ e menores que $s_{i+1}$.
 Os elementos em $f_t$ são todos maiores que $s_t$. Veja Figura~\ref{fig_noArvoreB}.

 Para encontrar uma chave $k \notin S$ da árvore, é preciso determinar em qual filho do nó $X$  continuar  pesquisa. Se $k < s_1$,  a procura continua no nó filho $f_0$. Se $ k > s_t$, a procura continua no nó filho $f_t$. Se $s_i < k < s_{i+1}$, a procura continua no filho $f_i$, entre as chaves $s_i$ e $s_{i+1}$ de $S$.

Os tempos das operações de uma árvore B estão diretamente ligadas à sua altura. O seguinte lema baseia-se em~\cite{cormen}.

\begin{lema}
Uma árvore B de grau $B \geq 4$ e altura $h$ respeita a seguinte relação:

\[ h =O (\log_B n) \]

\end{lema}

Para buscar uma chave $k$ em uma árvore B, deve-se, no pior caso, procurar sequencialmente $B - 1$ chaves em cada nó e repetir o processo em cada nível. Ou seja, o tempo de pesquisa de uma árvore B é de $O(B \log_B {n})$. Supondo que $B$ é constante, a pesquisa se dá em $\Theta (\lg n)$.

\begin{figure}[htb]
	\begin{center}
\begin{picture}(81,25)(0,0)
	\node[Nframe=n,Nadjust=wh,Nmr=1](i6)(6,9){} 
	\node[Nframe=n,Nadjust=wh,Nmr=1](f6)(6,1){\tiny$T_1$} 
	\drawedge[ATnb=0,AHnb=1](i6,f6){} 
	\node[Nframe=n,Nadjust=wh,Nmr=1](i1)(17,19.1){} 
	\node[Nframe=n,Nadjust=wh,Nmr=1](f1)(17,11.2){} 
	\drawedge[ATnb=0,AHnb=1](i1,f1){} 
	\node[Nframe=n,Nadjust=wh,Nmr=1](i20)(63.5,18.8){} 
	\node[Nframe=n,Nadjust=wh,Nmr=1](f20)(58,11.2){} 
	\drawedge[ATnb=0,AHnb=1](i20,f20){} 
	\node[Nframe=n,Nadjust=wh,Nmr=1](i21)(65.5,18.8){} 
	\node[Nframe=n,Nadjust=wh,Nmr=1](f21)(72,11.1){} 
	\drawedge[ATnb=0,AHnb=1](i21,f21){} 	
	\node[Nframe=n,Nadjust=wh,Nmr=1](i7)(21.5,8.7){} 
	\node[Nframe=n,Nadjust=wh,Nmr=1](f7)(21.5,1){\tiny$T_6$} 
	\drawedge[ATnb=0,AHnb=1](i7,f7){} 
	\node[Nframe=n,Nadjust=wh,Nmr=1](i8)(25,8.7){} 
	\node[Nframe=n,Nadjust=wh,Nmr=1](f8)(25,1){\tiny$T_7$} 
	\drawedge[ATnb=0,AHnb=1](i8,f8){} 
	\node[Nframe=n,Nadjust=wh,Nmr=1](i9)(28,9){} 
	\node[Nframe=n,Nadjust=wh,Nmr=1](f9)(28,1){\tiny$T_8$} 
	\drawedge[ATnb=0,AHnb=1](i9,f9){} 
	\node[Nframe=n,Nadjust=wh,Nmr=1](i2)(18.5,8.7){} 
	\node[Nframe=n,Nadjust=wh,Nmr=1](f2)(18.5,1){\tiny$T_5$} 
	\drawedge[ATnb=0,AHnb=1](i2,f2){} 
	\node[Nframe=n,Nadjust=wh,Nmr=1](i3)(15.5,8.7){} 
	\node[Nframe=n,Nadjust=wh,Nmr=1](f3)(15.5,1){\tiny$T_4$} 
	\drawedge[ATnb=0,AHnb=1](i3,f3){} 
	\node[Nframe=n,Nadjust=wh,Nmr=1](i4)(12.5,8.7){} 
	\node[Nframe=n,Nadjust=wh,Nmr=1](f4)(12.5,1){\tiny$T_3$} 
	\drawedge[ATnb=0,AHnb=1](i4,f4){} 
	\node[Nframe=n,Nadjust=wh,Nmr=1](i5)(9,8.7){} 
	\node[Nframe=n,Nadjust=wh,Nmr=1](f5)(9,1){\tiny$T_2$} 
		\drawedge[ATnb=0,AHnb=1](i5,f5){} 
	\node[Nframe=n,Nadjust=wh,Nmr=1](i17)(53.5,8.7){} 
	\node[Nframe=n,Nadjust=wh,Nmr=1](f17)(53.5,1){\tiny$T_1$} 
	\drawedge[ATnb=0,AHnb=1](i17,f17){} 
	\node[Nframe=n,Nadjust=wh,Nmr=1](i18)(56.5,8.7){} 
	\node[Nframe=n,Nadjust=wh,Nmr=1](f18)(56.5,1){\tiny$T_2$} 
	\drawedge[ATnb=0,AHnb=1](i18,f18){} 
	\node[Nframe=n,Nadjust=wh,Nmr=1](i19)(60,8.7){} 
	\node[Nframe=n,Nadjust=wh,Nmr=1](f19)(60,1){\tiny$T_3$} 
	\drawedge[ATnb=0,AHnb=1](i19,f19){} 
	\node[Nframe=n,Nadjust=wh,Nmr=1](i12)(63,9){} 
	\node[Nframe=n,Nadjust=wh,Nmr=1](f12)(63,1){\tiny$T_4$} 
	\drawedge[ATnb=0,AHnb=1](i12,f12){} 
	\node[Nframe=n,Nadjust=wh,Nmr=1](i13)(67,9.2){} 
	\node[Nframe=n,Nadjust=wh,Nmr=1](f13)(67,1){\tiny$T_5$} 
	\drawedge[ATnb=0,AHnb=1](i13,f13){} 
	\node[Nframe=n,Nadjust=wh,Nmr=1](i14)(70.4,9){} 
	\node[Nframe=n,Nadjust=wh,Nmr=1](f14)(70.4,1){\tiny$T_6$} 
	\drawedge[ATnb=0,AHnb=1](i14,f14){} 
	\node[Nframe=n,Nadjust=wh,Nmr=1](i15)(74,9){} 
	\node[Nframe=n,Nadjust=wh,Nmr=1](f15)(74,1){\tiny$T_7$} 	
	\drawedge[ATnb=0,AHnb=1](i15,f15){} 
	\node[Nframe=n,Nadjust=wh,Nmr=1](i16)(77,9.3){} 
	\node[Nframe=n,Nadjust=wh,Nmr=1](f16)(77,1){\tiny$T_8$} 	
	\drawedge[ATnb=0,AHnb=1](i16,f16){} 		
		\node[Nframe=n,Nadjust=wh](x)(30,15){} 	
		\node[Nframe=n,Nadjust=wh](y)(50,15){} 	
		\drawedge[ATnb=0,AHnb=1,linewidth=.4,AHLength=2.5](x,y){Separação} 	
		\drawedge[ELside=r](x,y){$O(B)$} 			
	\node[Nadjust=wh,Nmr=1](A1)(17,20){\footnotesize $\ldots$ N W $\ldots$} 
	\node[Nadjust=wh,Nmr=1](B1)(17,10){\footnotesize P Q R {\bf S} T U V} 
	\node[Nadjust=wh,Nmr=1](A2)(65,20){\footnotesize $\ldots$ N {\bf S} W $\ldots$} 
	\node[Nadjust=wh,Nmr=1](A)(58,10){\footnotesize P Q R} 
	\node[Nadjust=wh,Nmr=1](A)(72,10){\footnotesize T U V} 
	\end{picture}	

	\caption{Inserção de uma chave em um nó completo da árvore B. Fonte:~\cite{cormen}.}
	\label{fig_separaArvoreB}
	\end{center}
\end{figure}
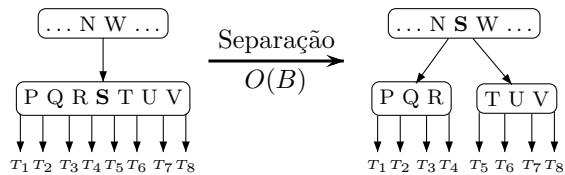

Para se inserir uma chave $k$ na árvore B, primeiro deve pesquisar o nó em que a nova chave será inserida. 
Se houver espaço no nó, gasta-se mais $O(B)$
para acomodar a chave no nó. Este é o custo para inserir um elemento em uma posição central de um vetor de tamanho B.
Se o nó que receberá a chave estiver lotado, ele deve ser dividido ao meio. Veja Figura~\ref{fig_separaArvoreB}.
Essa separação acontece da seguinte maneira: a chave mediana $s_m$ do nó em questão vai para o nó pai. São criados dois nós, um com as chaves $s_i <  s_m$ e outro com as chaves $s_i > s_m$. Tais nós passam a ser filhos à esquerda e à direita de $s_m$, cada um com exatamente $\frac{B}{2}-1$ chaves.

O custo de dividir um nó ao meio é $O(B)$ através de operações elementares em vetores.
 Pode-se perceber que o nó pai pode ser um nó completo também. Neste caso, o processo de separação se propaga para cima da árvore, possivelmente até a raiz.

Se esse processo se propagar até a raiz da árvore  a separação será realizada em tempo $O(B h) = O(B \cdot \log_B n)$. 
Esta complexidade de pior caso, entretanto, pode melhorar através de uma análise amortizada. Cada separação custa $O(B)$ e cria um novo nó. Como, 
ao final da iserção de $n$ elementos teremos no máximo a  $1+\frac{n-1}{B/2-1}$ nós, o custo de todas as separações será $O(B(1+\frac{n-1}{B/2-1}))=O(n)$.
Ou seja, após a inserção de $n$ elementos, será gasto $O(n)$ para realizar todas as separações.

Removendo-se o custo da separação, cada inserção levará $O(B\log_B n+B)$ que é o custo para se encontrar o nó mais o custo para se inserir uma chave em um nó.

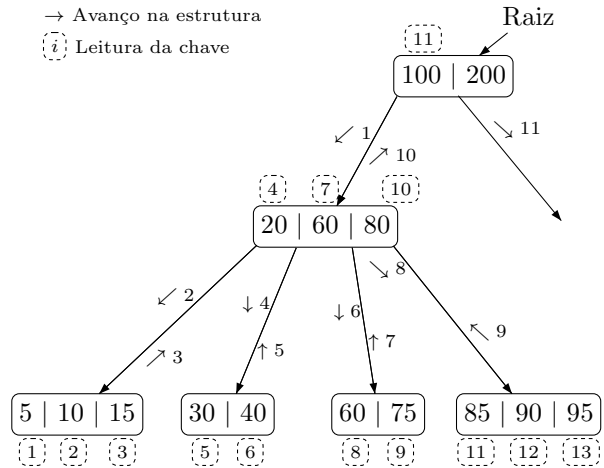
\begin{figure}[h!]
	\begin{center}
	\begin{picture}(80,60)(0,0)
	\node[Nframe=n,Nadjust=wh,Nmr=1](R)(70,58){Raiz} 
	\node[Nadjust=wh,Nmr=1](n1)(60,50){$100~|~200$} 
	\node[Nframe=n,Nadjust=wh,Nmr=1](i1)(34.6,28.2){} 
	\node[Nframe=n,Nadjust=wh,Nmr=1](i2)(39.5,28.1){} 
	\node[Nframe=n,Nadjust=wh,Nmr=1](i3)(46.4,28.2){} 
	\node[Nframe=n,Nadjust=wh,Nmr=1](i4)(51.4,28.2){} 
	\node[Nframe=n,Nadjust=wh,Nmr=1](i5)(53,48.2){} 	
	\node[Nframe=n,Nadjust=wh,Nmr=1](i6)(60,48.2){} 	
	\node[dash={.5}0,Nadjust=wh,Nmr=1](x1)(7,54){\scriptsize$i$} 
	\node[dash={.5}0,Nadjust=wh,Nmr=1](x1)(4,0){\scriptsize$1$} 
	\node[dash={.5}0,Nadjust=wh,Nmr=1](x1)(9.5,0){\scriptsize$2$} 
	\node[dash={.5}0,Nadjust=wh,Nmr=1](x1)(16,0){\scriptsize$3$} 	
	\node[dash={.5}0,Nadjust=wh,Nmr=1](x1)(27,0){\scriptsize$5$} 	
	\node[dash={.5}0,Nadjust=wh,Nmr=1](x1)(33,0){\scriptsize$6$} 	
	\node[dash={.5}0,Nadjust=wh,Nmr=1](x1)(47,0){\scriptsize$8$} 	
	\node[dash={.5}0,Nadjust=wh,Nmr=1](x1)(53,0){\scriptsize$9$} 			
	\node[dash={.5}0,Nadjust=wh,Nmr=1](x1)(63,0){\scriptsize$11$}
	\node[dash={.5}0,Nadjust=wh,Nmr=1](x1)(70,0){\scriptsize$12$}
	\node[dash={.5}0,Nadjust=wh,Nmr=1](x1)(77,0){\scriptsize$13$}
	\node[dash={.5}0,Nadjust=wh,Nmr=1](x1)(36,35){\scriptsize$4$} 	
	\node[dash={.5}0,Nadjust=wh,Nmr=1](x1)(43,35){\scriptsize$7$} 	
	\node[dash={.5}0,Nadjust=wh,Nmr=1](x1)(53,35){\scriptsize$10$} 		
	\node[dash={.5}0,Nadjust=wh,Nmr=1](x1)(56,55){\scriptsize$11$} 
			
	\node[Nframe=n,Nadjust=wh,Nmr=1](X)(20,58){\scriptsize$\rightarrow$ Avanço na estrutura} 	
	\node[Nframe=n,Nadjust=wh,Nmr=1](X)(20,54){\scriptsize ~~~~ Leitura da chave~~~~~} 
	\drawedge[ATnb=0,AHnb=1](R,n1){} 
	\node[Nadjust=wh,Nmr=1](n2)(43,30){$20~|~60~|~80$} 
	\node[Nframe=n,Nadjust=wh,Nmr=1](n3)(75,30){} 
	\drawedge[ATnb=0,AHnb=1,ELpos=40,ELside=r,ELdist=0](i5,n2){\scriptsize$\swarrow 1$} 
	\drawedge[ATnb=0,AHnb=1,ELpos=40,ELdist=0](i5,n2){\scriptsize$\nearrow 10$} 
	\drawedge[ATnb=0,AHnb=1,ELpos=40,ELdist=0](i6,n3){\scriptsize$\searrow 11$} 
	\node[Nadjust=wh,Nmr=1](n4)(10,5){$5~|~10~|~15$} 
	\node[Nadjust=wh,Nmr=1](n6)(30,5){$30~|~40$} 
	\node[Nadjust=wh,Nmr=1](n7)(50,5){$60~|~75$} 
	\node[Nadjust=wh,Nmr=1](n9)(70,5){$85~|~90~|~95$} 
	
	\drawedge[ATnb=0,AHnb=1,ELpos=40,ELside=r,ELdist=0](i1,n4){\scriptsize$\swarrow 2$}
	\drawedge[ATnb=0,AHnb=0,ELpos=40,ELside=r,ELdist=0](n4,i1){\scriptsize$\nearrow 3$}

	\drawedge[ATnb=0,AHnb=1,ELpos=40,ELside=r,ELdist=0](i2,n6){\scriptsize$\downarrow 4$}
	\drawedge[ATnb=0,AHnb=0,ELpos=40,ELside=r,ELdist=0](n6,i2){\scriptsize$\uparrow 5$}

	\drawedge[ATnb=0,AHnb=1,ELpos=40,ELside=r,ELdist=0](i3,n7){\scriptsize$\downarrow 6$}
	\drawedge[ATnb=0,AHnb=0,ELpos=40,ELside=r,ELdist=0](n7,i3){\scriptsize$\uparrow 7$}

	\drawedge[ATnb=0,AHnb=1,ELpos=10,ELside=r,ELdist=0](i4,n9){\scriptsize$\searrow 8$}
	\drawedge[ATnb=0,AHnb=0,ELpos=40,ELside=r,ELdist=0](n9,i4){\scriptsize$\nwarrow 9$}

	\end{picture}	
	\caption{Visitação em ordem de uma árvore B}
	\label{fig_pesquisaOrdenadaArvoreB}
	\end{center}
\end{figure}

Para ordenar uma sequência de $n$ números inteiros utilizando uma árvore B, é necessário  inserir todos os elementos em uma árvore inicialmente vazia. Ao final, tem-se um árvore B com os $n$ elementos da sequência inicial. Para se obter a sequência ordenada, é realizado um percurso em ordem na árvore, como demonstrado na Figura~\ref{fig_pesquisaOrdenadaArvoreB}. As setas contínuas representam para onde caminhar na árvore, e as setas pontilhadas representam a leitura da chave na sequência. Cada seta possui um número, que indica a sequência dos elementos visitados no percurso em ordem.

O tempo para ordenarmos $n$ números será então a soma dos tempos de inserção das $n$ chaves, ou seja, $O(n \cdot B \log_B n+nB)$. 
Se B for constante, a complexidade final será $O(n \lg n)$.

\section{Árvores de Fusão}

Nesta seção será descrita a estrutura Árvore de Fusão proposta por~\cite{fredman}. Uma Árvore de Fusão é semelhante a uma árvore B. Seja $t$  o grau em um nó de uma árvore B.
Considerando-se um nó não raiz e não folha, o valor de  $t$ está entre $\frac{B}{2}$ e B.  O mesmo vale para a estrutura de dados
Árvore de Fusão. Entretando,  o valor de $B$  na Árvore de Fusão não é constante mas sim uma função de $n$. Mais precisamente $B=(\lg n)^\frac{1}{5}$. 
Outra diferença entre árvore B e Árvore de Fusão é que na primeira  gasta-se $B$ para fazer a busca de uma chave $k$
dentro de um nó, enquanto numa Árvore de Fusão, a busca por uma chave é feita em $O(1)$.

Considere o problema do sucessor/predecessor de uma chave $x$ em um conjunto $S$. Este problema
consiste em encontrar o número imediatamente inferior/superior a $x$ em $S$.
Árvores de Fusão são estruturas de dados semelhantes  a árvores B, 
mas que conseguem resolver o problema predecessor/sucessor em tempo $O(1)$ dentro de um nó.
 Dada uma chave de pesquisa $x$, a Árvore de Fusão consegue achar
o ramo filho relativo a $x$ em tempo constante, apesar de possuir um número de filhos crescente em relação a $n$.

A notação a seguir é necessária para o estudo de Árvore de Fusão~\cite{fredman}:

\begin{definicao}{$\rk{x}$:}
Dado um conjunto $S$ de números inteiros e um número $x$, denota-se $\rk{x}$ o valor $|\{t~ |~ t\in S, t \leq x\}|$. Em outras palavras, $\rk{x}$ representa a quantidade de números em $S$ menores ou iguais a $x$. 
\end{definicao}

Veja que o problema de ordenar $n$ números pode ser resumido a encontrar $\rk{x}$, pois essa função fornece a posição exata de $x$ no vetor ordenado. 
Além disso,  $\rk{x}$ indica o nó filho para continuar a busca de um elemento $x$ em um nó de uma árvore B.
Como a Árvore de Fusão baseia-se na estrutura de dados $\tr$ do trabalho~\cite{ajtai}, tal estrutura será descrita a seguir.

\subsection{Estrutura de Dados $\tr$ de~\cite{ajtai}}

Seja uma $\tr$ uma árvore Binária construída da seguinte forma. Dado um número binário  $x$  com $w$ bits,
cada bit de $x$ é um nó na $\tr$.
Se o bit mais significativo de $x$ for zero, $x$ é um filho à esquerda da raiz. Se for 1, $x$ é um filho à direita da raiz. Esta propriedade vale para os filhos, a partir do próximo bit. 

\begin{figure}[htb]
	\begin{center}
	\setlength{\unitlength}{.8mm}
	\begin{picture}(70,90)(0,0)
	\node[Nadjust=wh,Nmr=3](b0)(10,85){\small $b_7$} 
	\node[Nadjust=wh,Nmr=3](b1)(20,75){\small $b_6$} 
	\node[Nadjust=wh,Nmr=3](b2)(30,65){\small $b_5$} 
	\node[Nadjust=wh,Nmr=3,linewidth=0.5](b3)(40,55){\small $b_4$} 
	\node[Nadjust=wh,Nmr=3](b4)(50,45){\small $b_3$} 
	\node[Nadjust=wh,Nmr=3](c4)(30,45){\small $b_3$} 
	\node[Nadjust=wh,Nmr=3](b5)(60,35){\small $b_2$} 
	\node[Nadjust=wh,Nmr=3](c5)(40,35){\small $b_2$} 
	\node[Nadjust=wh,Nmr=3](b6)(50,25){\small $b_1$} 
	\node[Nadjust=wh,Nmr=3](c6)(30,25){\small $b_1$} 
	\node[Nadjust=wh,Nmr=3](b7)(40,15){\small $b_0$} 
	\node[Nadjust=wh,Nmr=3](c7)(20,15){\small $b_0$} 
	\node[Nadjust=wh,Nmr=0](c8)(30,5){\small $s_1$} 
	\node[Nadjust=wh,Nmr=0](b8)(50,5){\small $s_2$} 
	
	\drawedge[ATnb=0,AHnb=1](b0,b1){\small 1} 
	\drawedge[ATnb=0,AHnb=1](b1,b2){\small 1}  
	\drawedge[ATnb=0,AHnb=1](b2,b3){\small 1}
	\drawedge[ATnb=0,AHnb=1](b3,b4){\small 1}
	\drawedge[ATnb=0,AHnb=1](b4,b5){\small 1} 
	\drawedge[ATnb=0,AHnb=1](b5,b6){\small 0}   
	\drawedge[ATnb=0,AHnb=1](b6,b7){\small 0} 
	\drawedge[ATnb=0,AHnb=1](b7,b8){\small 1} 
	\drawedge[ATnb=0,AHnb=1](b3,c4){\small 0}
	\drawedge[ATnb=0,AHnb=1](c4,c5){\small 1} 
	\drawedge[ATnb=0,AHnb=1](c5,c6){\small 0}   
	\drawedge[ATnb=0,AHnb=1](c6,c7){\small 0} 
	\drawedge[ATnb=0,AHnb=1](c7,c8){\small 1}

	\end{picture}	
	\setlength{\unitlength}{1mm}
	\caption{Estrutura $\tr$ para $s_1$ e $s_2$. Basta analisar o $b_4$ para ordená-los}
	\label{fig_trieBit}
	\end{center}
\end{figure}
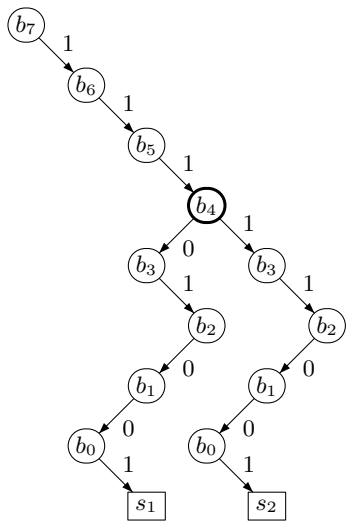

Dado um inteiro qualquer, seja  $b_i$  seu $i$-ésimo bit menos significativo. Assim $b_0$ é o bit menos significativo, $b_1$ o  segundo bit
menos significativo e assim por diante. 
Considere dois números binários $s_1 = 11101001$  e $s_2 = 11111001$.
 A Figura~\ref{fig_trieBit} apresenta uma $\tr$ contendo $s_1$ e $s_2$.
 Note que as folhas da $\tr$ estão sempre ordenadas. Suponha também que 
 o $\rk{x}$ está calculado para todos os elementos da $\tr$.

Considere a seguinte definição:
\begin{definicao}{$\dlt{s_1}{s_2}$:}
Dados dois inteiros $s_1$ e $s_2$, seja $\dlt{s_1}{s_2}$ o bit de interesse entre $s_1$ e $s_2$,
ou seja, o bit mais significativo que $s_1$ diverge de   $s_2$.
\end{definicao}

\begin{figure}[htb]
	\begin{center}
	\setlength{\unitlength}{.8mm}
	\begin{picture}(70,30)(0,0)
	\node[Nadjust=wh,Nmr=3](b3)(35,20){\small $b_4$} 
	\node[Nframe=n,Nadjust=wh,Nmr=3](x)(65,20){\small bit de interesse} 
	
	\node[Nadjust=wh,Nmr=0](c8)(25,5){\small $s_1$} 
	\node[Nadjust=wh,Nmr=0](b8)(45,5){\small $s_2$} 
	\drawedge[ATnb=0,AHnb=1](b3,b8){\small 1}
	\drawedge[ATnb=0,AHnb=1,ELside=r](b3,c8){\small 0}
	\drawedge[ATnb=0,AHnb=1](x,b3){}
	\end{picture}	
	\setlength{\unitlength}{1mm}
	\caption{Estrutura $\trc$ para comparar $s_1$ e $s_2$ com somente o bit de interesse}
	\label{fig_trieBitCondensado}
	\end{center}
\end{figure}
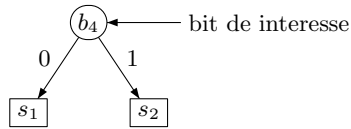

Para comparar números binários, não é necessário comparar todos os bits, e sim os \textbf{bits de interesse}, que são os bits mais significativos que diferenciam um conjunto de números. Por exemplo, para compararmos os números binários $s_1=11101001$ e $s_2=11111001$, basta comparar o bit mais significativo que diverge em $s_1$ e $s_2$. Nesse exemplo, o bit que se deve comparar é o $b_4$, que em $s_1$ é 0 e em $s_2$ vale 1. Sendo assim, $\dlt{s_1}{s_2}=b_4$.
A partir disso, se sabe que o número $s_2$ é maior que $s_1$ sem analisar o restante dos bits. Veja Figura~\ref{fig_trieBitCondensado}.

Seja $\oplus$ um xor bit a bit entre duas palavras.
Dados dois inteiros $s_1$ e $s_2$, $\dlt{s_1}{s_2}$ pode ser obtido pela seguinte fórmula: 
$$\dlt{s_1}{s_2}=\lfloor\lg (s_1 \oplus  s_2)\rfloor.$$

Considere uma sequência de inteiros $S = (s_1, \ldots , s_n)$ e uma estrutura de dados $\tr$. Após inserir todos
os elementos de $S$ na $\tr$, será feita uma compressão  onde serão guardados apenas os bits de interesse.
Esta nova árvore será chamada de $\trc$

A $\trc$ guarda cada elemento de $S$ em uma folha. Já os nós internos da $\trc$ guardam o bit de interesse para comparação entre seus dois filhos.

\begin{lema}
Dada uma $\trc$ contendo $S=(s_1,\ldots,s_t)$, o número de bits de interesse será menor igual a $t-1$.  
\label{nbits}
\end{lema}

O Lema~\ref{nbits} decorre do fato de cada bit de interesse criar uma ramificação na $\tr$. O número de ramificações
será exatamente $t-1$. Eventualmente, duas ramificações distintas  podem ocorrer 
no mesmo bit de interesse.

 Para buscar $x$ em uma $\trc$, são comparados os bits de $x$  com os bits da $\tr$ da raiz até a folha. 
  Em cada nó, caso o bit de $x$ seja 0, a busca continua no filho à esquerda. Caso o bit seja 1, a busca continua no filho à direita. O respectivo bit considerado em cada nível está armazenado no nó da $\trc$
  A Figura~\ref{fig_pesquisaTrie} ilustra uma pesquisa com o elemento $x$ em uma $\trc$ contendo os elementos $a, b, c $ e $d$. 
  Seja $\trs{x}$ o resultado desta busca. Na  Figura~\ref{fig_pesquisaTrie}, $\trs{x}=c$

Ao realizar esse procedimento, chega-se a uma folha, que é um elemento $s$ de $S$. Isso  mostra que $x$ e $s$ têm os mesmos bits nas posições percorridas no caminho da $\tr$.

  No exemplo da Figura~\ref{fig_pesquisaTrie} nota-se que o primeiro bit a divergir entre $x$ e $c$ é o $b_2$,
  ou seja $\dlt{x}{c}=b_2$.
  A Figura~\ref{fig_insercaoTrie} mostra como a $\tr$ fica após a inserção do elemento $x$ na estrutura.

\begin{figure*}[htb]
	\begin{center}
\begin{picture}(120,80)(0,0)
	\put(70,60){\begin{tabular}{cc}
	  S=\{a,b,c,d\}~~~~\\
	  a = 1 1 {\bf 0 1} 1 1 1 {\bf 1}\\
	  b = 1 1 {\bf 1 0}  0 0 0 {\bf 0}\\
	  c = 1 1 {\bf 1 0} 0 0 0 {\bf 1}\\
	  d = 1 1 {\bf 1 1} 1 1 1 {\bf 0}\\
	~\\		
	 x = 1 1 {\bf 1 0} 0 1 1 {\bf 1}			
	\end{tabular}}
	\node[Nadjust=wh,Nmr=3](fb2)(85,35){\small $b_5$} 
	\node[Nadjust=wh,Nmr=3](fb3)(95,25){\small $b_4$} 
	\node[Nadjust=wh,Nmr=3](fb7)(85,15){\small $b_0$} 
	\node[Nadjust=wh,Nmr=0](fc)(105,15){\small $d$} 
	\node[Nadjust=wh,Nmr=0](fb)(75,5){\small $b$} 
	\node[Nadjust=wh,Nmr=0,linewidth=.4](fd)(95,5){\small $c$} 
	\node[Nframe=n,Nadjust=wh,Nmr=0](x)(100,5){\small $\mathbf{x}$} 
	\node[Nadjust=wh,Nmr=0](fa)(75,25){\small $a$} 
	\drawedge[ATnb=0,AHnb=1,ELside=r,linewidth=.5,AHLength=2,ELpos=70](fb3,fb7){\small $x[4]=0$}
	\drawedge[ATnb=0,AHnb=1,ELside=l](fb3,fc){\small 1}
	\drawedge[ATnb=0,AHnb=1,ELside=l,linewidth=.5,AHLength=2,ELpos=70](fb7,fd){\small $x[0]=1$}
	\drawedge[ATnb=0,AHnb=1,ELside=r](fb7,fb){\small 0}
	\drawedge[ATnb=0,AHnb=1,ELside=l,linewidth=.5,AHLength=2,ELpos=70](fb2,fb3){\small $x[5]=1$}
	\drawedge[ATnb=0,AHnb=1,ELside=r](fb2,fa){\small 0}	
	
		\node[Nadjust=wh,Nmr=3](b0)(10,75){\small $b_7$} 
	\node[Nadjust=wh,Nmr=3](b1)(20,70){\small $b_6$} 
	\node[Nadjust=wh,Nmr=3,linewidth=0.5](b2)(30,65){\small $b_5$} 
	\node[Nadjust=wh,Nmr=3,linewidth=0.5](b3)(40,55){\small $b_4$} 
	\node[Nadjust=wh,Nmr=3](a3)(00,55){\small $b_4$} 
	\node[Nadjust=wh,Nmr=3](a4)(3,45){\small $b_3$} 
	\node[Nadjust=wh,Nmr=3](a5)(6,35){\small $b_2$} 
	\node[Nadjust=wh,Nmr=3](a6)(9,25){\small $b_1$} 
	\node[Nadjust=wh,Nmr=3](a7)(12,15){\small $b_0$} 
	\node[Nadjust=wh,Nmr=0](a8)(15,5){\small a} 
	\drawedge[ATnb=0,AHnb=1,ELside=r](b2,a3){\small 0} 
	\drawedge[ATnb=0,AHnb=1](a3,a4){\small 1} 
	\drawedge[ATnb=0,AHnb=1](a4,a5){\small 1} 	
	\drawedge[ATnb=0,AHnb=1](a5,a6){\small 1} 	
	\drawedge[ATnb=0,AHnb=1](a6,a7){\small 1} 	
	\drawedge[ATnb=0,AHnb=1](a7,a8){\small 1} 
	
	\node[Nadjust=wh,Nmr=3](b4)(47,45){\small $b_3$} 
	\node[Nadjust=wh,Nmr=3](b5)(50,35){\small $b_2$} 
	\node[Nadjust=wh,Nmr=3](b6)(53,25){\small $b_1$} 
	\node[Nadjust=wh,Nmr=3](b7)(56,15){\small $b_0$} 
	\node[Nadjust=wh,Nmr=0](b8)(53,5){\small d} 
	
	\node[Nadjust=wh,Nmr=3](c4)(37,45){\small $b_3$}

	\node[Nadjust=wh,Nmr=3](c5)(34,35){\small $b_2$} 
	\node[Nadjust=wh,Nmr=3](c6)(31,25){\small $b_1$} 
	\node[Nadjust=wh,Nmr=3,linewidth=0.5](c7)(28,15){\small $b_0$} 
	\node[Nadjust=wh,Nmr=0](c8)(25,5){\small b} 
	\node[Nadjust=wh,Nmr=0](c9)(31,5){\small c}

	\drawedge[ATnb=0,AHnb=1](b0,b1){\small 1} 
	\drawedge[ATnb=0,AHnb=1](b1,b2){\small 1}  
	\drawedge[ATnb=0,AHnb=1](b2,b3){\small 1}
	\drawedge[ATnb=0,AHnb=1](b3,b4){\small 1}
	\drawedge[ATnb=0,AHnb=1](b4,b5){\small 1} 
	\drawedge[ATnb=0,AHnb=1](b5,b6){\small 1}   
	\drawedge[ATnb=0,AHnb=1](b6,b7){\small 1} 
	\drawedge[ATnb=0,AHnb=1](b7,b8){\small 0} 
	\drawedge[ATnb=0,AHnb=1,ELside=r](b3,c4){\small 0}
	\drawedge[ATnb=0,AHnb=1,ELside=r](c4,c5){\small 0} 
	\drawedge[ATnb=0,AHnb=1,ELside=r](c5,c6){\small 0}   
	\drawedge[ATnb=0,AHnb=1,ELside=r](c6,c7){\small 0} 
	\drawedge[ATnb=0,AHnb=1,ELside=r](c7,c8){\small 0} 
	\drawedge[ATnb=0,AHnb=1,ELside=l](c7,c9){\small 1} 	
	
	\end{picture}	
	\caption{Pesquisa de uma chave $x$ na estrutura de dados $\trc$.}
	\label{fig_pesquisaTrie}
	\end{center}
\end{figure*}
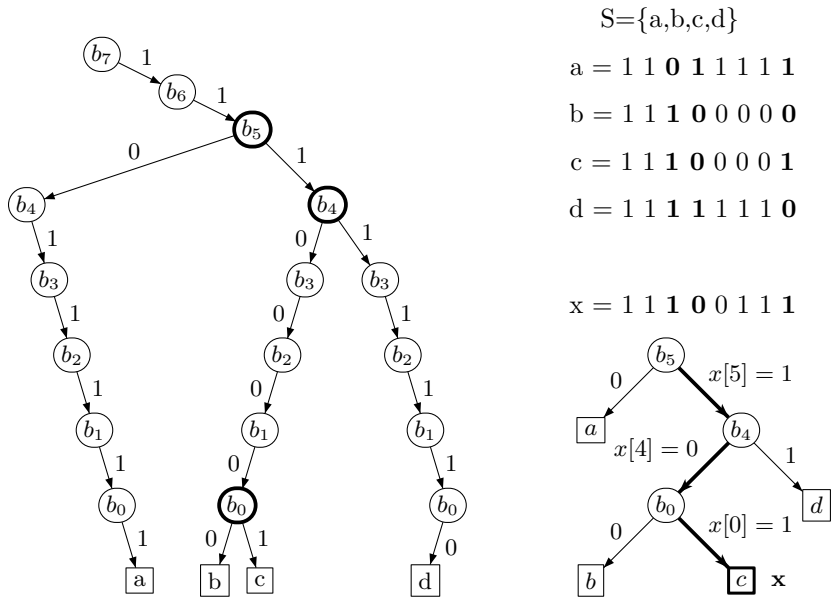

\begin{figure*}[htb]
	\begin{center}
	\setlength{\unitlength}{.9mm}
\begin{picture}(120,80)(0,0)
	\put(70,65){\begin{tabular}{cc}
	  S=\{a,b,c,d\}~~~~\\
	  a = 1 1 {\bf 0 1} 1 1 1 {\bf 1}\\
	  b = 1 1 {\bf 1 0}  0 0 0 {\bf 0}\\
	  c = 1 1 {\bf 1 0} 0 0 0 {\bf 1}\\
	  d = 1 1 {\bf 1 1} 1 1 1 {\bf 0}\\
	 x = 1 1 {\bf 1 0} 0 1 1 {\bf 1}			
	\end{tabular}}
	\node[Nadjust=wh,Nmr=3](fb2)(85,45){\small $b_5$} 
	\node[Nadjust=wh,Nmr=3](fb3)(93,35){\small $b_4$} 
	\node[Nadjust=wh,Nmr=3,linewidth=0.5](fb5)(85,25){\small $b_2$} 
	\node[Nadjust=wh,Nmr=3](fb7)(77,15){\small $b_0$} 
	\node[Nadjust=wh,Nmr=0](fc)(99,25){\small d} 
	\node[Nadjust=wh,Nmr=0](fb)(71,5){\small b} 
	\node[Nadjust=wh,Nmr=0](fd)(85,5){\small c} 
	\node[Nadjust=wh,Nmr=0](x)(93,15){\small x} 
	\node[Nadjust=wh,Nmr=0](fa)(77,35){\small $a$} 
	\drawedge[ATnb=0,AHnb=1,ELside=r](fb3,fb5){\small $0$}
	\drawedge[ATnb=0,AHnb=1,ELside=r](fb5,fb7){\small $0$}
	\drawedge[ATnb=0,AHnb=1,ELside=l](fb3,fc){\small 1}
	\drawedge[ATnb=0,AHnb=1,ELside=l](fb5,x){\small 1}
	\drawedge[ATnb=0,AHnb=1,ELside=l](fb7,fd){\small $1$}
	\drawedge[ATnb=0,AHnb=1,ELside=r](fb7,fb){\small 0}
	\drawedge[ATnb=0,AHnb=1,ELside=l](fb2,fb3){\small $1$}
	\drawedge[ATnb=0,AHnb=1,ELside=r](fb2,fa){\small 0}	
	
		\node[Nadjust=wh,Nmr=3](b0)(10,75){\small $b_7$} 
	\node[Nadjust=wh,Nmr=3](b1)(20,70){\small $b_6$} 
	\node[Nadjust=wh,Nmr=3](b2)(30,65){\small $b_5$} 
	\node[Nadjust=wh,Nmr=3](b3)(40,55){\small $b_4$} 
	\node[Nadjust=wh,Nmr=3](a3)(00,55){\small $b_4$} 
	\node[Nadjust=wh,Nmr=3](a4)(3,45){\small $b_3$} 
	\node[Nadjust=wh,Nmr=3](a5)(6,35){\small $b_2$} 
	\node[Nadjust=wh,Nmr=3](a6)(9,25){\small $b_1$} 
	\node[Nadjust=wh,Nmr=3](a7)(12,15){\small $b_0$} 
	\node[Nadjust=wh,Nmr=0](a8)(15,5){\small a} 
	\drawedge[ATnb=0,AHnb=1,ELside=r](b2,a3){\small 0} 
	\drawedge[ATnb=0,AHnb=1](a3,a4){\small 1} 
	\drawedge[ATnb=0,AHnb=1](a4,a5){\small 1} 	
	\drawedge[ATnb=0,AHnb=1](a5,a6){\small 1} 	
	\drawedge[ATnb=0,AHnb=1](a6,a7){\small 1} 	
	\drawedge[ATnb=0,AHnb=1](a7,a8){\small 1} 
	
	\node[Nadjust=wh,Nmr=3](b4)(47,45){\small $b_3$} 
	\node[Nadjust=wh,Nmr=3](b5)(50,35){\small $b_2$} 
	\node[Nadjust=wh,Nmr=3](b6)(53,25){\small $b_1$} 
	\node[Nadjust=wh,Nmr=3](b7)(56,15){\small $b_0$} 
	\node[Nadjust=wh,Nmr=0](b8)(53,5){\small d} 
	
	\node[Nadjust=wh,Nmr=3](c4)(37,45){\small $b_3$}

	\node[Nadjust=wh,Nmr=3,linewidth=0.5](c5)(34,35){\small $b_2$} 
	\node[Nadjust=wh,Nmr=3](c6)(31,25){\small $b_1$} 
	\node[Nadjust=wh,Nmr=3](x6)(40,25){\small $b_1$} 
	\node[Nadjust=wh,Nmr=3](x7)(42,15){\small $b_0$} 
	\node[Nadjust=wh,Nmr=0](x8)(45,5){\small x} 
	\drawedge[ATnb=0,AHnb=1](c5,x6){\small 1} 
	\drawedge[ATnb=0,AHnb=1](x6,x7){\small 1} 
	\drawedge[ATnb=0,AHnb=1](x7,x8){\small 1} 			

	\node[Nadjust=wh,Nmr=3](c7)(28,15){\small $b_0$} 
	\node[Nadjust=wh,Nmr=0](c8)(25,5){\small b} 
	\node[Nadjust=wh,Nmr=0](c9)(31,5){\small c}

	\drawedge[ATnb=0,AHnb=1](b0,b1){\small 1} 
	\drawedge[ATnb=0,AHnb=1](b1,b2){\small 1}  
	\drawedge[ATnb=0,AHnb=1](b2,b3){\small 1}
	\drawedge[ATnb=0,AHnb=1](b3,b4){\small 1}
	\drawedge[ATnb=0,AHnb=1](b4,b5){\small 1} 
	\drawedge[ATnb=0,AHnb=1](b5,b6){\small 1}   
	\drawedge[ATnb=0,AHnb=1](b6,b7){\small 1} 
	\drawedge[ATnb=0,AHnb=1](b7,b8){\small 0} 
	\drawedge[ATnb=0,AHnb=1,ELside=r](b3,c4){\small 0}
	\drawedge[ATnb=0,AHnb=1,ELside=r](c4,c5){\small 0} 
	\drawedge[ATnb=0,AHnb=1,ELside=r](c5,c6){\small 0}   
	\drawedge[ATnb=0,AHnb=1,ELside=r](c6,c7){\small 0} 
	\drawedge[ATnb=0,AHnb=1,ELside=r](c7,c8){\small 0} 
	\drawedge[ATnb=0,AHnb=1,ELside=l](c7,c9){\small 1} 	
	
	\end{picture}	
	\setlength{\unitlength}{1mm}
	\caption{$\tr$  e $\trc$ após inserção do elemento $x$ \cite{ajtai}.}
	\label{fig_insercaoTrie}
	\end{center}
\end{figure*}

  \subsubsection{Calculando o $\rk{x}$}
  
  Suponha um conjunto $S=(s_1,\ldots,s_t)$ inseridos numa $\trc$. Esta seção mostrará como 
  calcular $\rk{x}$ para uma dada chave $x$. Primeiro, calcula-se $s'=\trs{x}$. 
  O elemento $s'$ possui valores igual a $x$ nos bits de interesse.
  Se $s'$ for igual a $x$ nos demais bits, $\rk{x}=\rk{s'}+1$ e o problema se encerra.
Caso contrário, será necessária uma segunda busca. Será calculado o bit de interesse $b'=\dlt{x}{s'}$. 

\begin{lema} O bit de interesse $b'=\dlt{x}{s'}$ é o novo vit de interesse da $\trc$ com $S\cup\{x\}$.
\end{lema}

Considere dois casos. No primeiro o bit $b'$ de $x$ vale $1$, ou seja $x[b']=1$ enquanto no segundo caso $x[b']=0$.

\begin{lema} 
Os bits mais significativos de $x$ e $s'$ são idênticos. O primeiro bit a divergir é $b'$. Considere 
a ramificação entre $x$ e $s'$ na $\tr$ contendo $S\cup\{x\}$. Se $x[b']=1$, o predecessor de 
$x$ é o maior elemento no  ramo $b'=0$. Se $x[b']=0$, o sucessor de $x$ é o menor elemento
no ramo $b'=1$.
\end{lema}

{\bf Caso  a ($x[b']=1$)} No exemplo da Figura~\ref{rk}, verifica-se que o predecessor de $x$ é o maior elemento da 
sub-árvore em destaque.

\begin{figure*}[htb]
	\begin{center}
	\setlength{\unitlength}{1mm}
\begin{picture}(160,80)(0,0)
	\put(128,2){\dashbox(15,17)}	
	\put(0,60){\begin{tabular}{r}
	  S=\{a,b,c,d\}\hspace{1.07cm}~\\
	  a = 1 1 {\bf 0 1} 1 1 1 {\bf 1}\\
	  b = 1 1 {\bf 1 0}  0 0 0 {\bf 0}\\
	  c = 1 1 {\bf 1 0} 0 0 0 {\bf 1}\\
	 d = 1 1 {\bf 1 1} 1 1 1 {\bf 0}\\
	~\\		
	 $x_1$ = 1 1 {\bf 1 0} 0 1 1 {\bf 1}			
	\end{tabular}}

	\node[Nframe=n,Nadjust=wh](x)(33,25){} 	
	\node[Nframe=n,Nadjust=wh](y)(53,25){} 	
	\drawedge[ATnb=0,AHnb=1,linewidth=.4,AHLength=2.5](x,y){$\dlt{x_1}{c}=b_2$} 	

	\node[Nadjust=wh,Nmr=3](fb2)(15,35){\small $b_5$} 
	\node[Nadjust=wh,Nmr=3](fb3)(25,25){\small $b_4$} 
	\node[Nadjust=wh,Nmr=3](fb7)(15,15){\small $b_0$} 
	\node[Nadjust=wh,Nmr=0](fc)(35,15){\small $d$} 
	\node[Nadjust=wh,Nmr=0](fb)(5,5){\small $b$} 
	\node[Nadjust=wh,Nmr=0,linewidth=.4](fd)(25,5){\small $c$} 
	\node[Nframe=n,Nadjust=wh,Nmr=0](x)(30,5){\small  $x_1$} 
	\node[Nadjust=wh,Nmr=0](fa)(5,25){\small $a$} 
	\drawedge[ATnb=0,AHnb=1,ELside=r,linewidth=.5,AHLength=2,ELpos=70](fb3,fb7){\small $x_1[4]=0$}
	\drawedge[ATnb=0,AHnb=1,ELside=l](fb3,fc){\small 1}
	\drawedge[ATnb=0,AHnb=1,ELside=l,linewidth=.5,AHLength=2,ELpos=70](fb7,fd){\small $x_1[0]=1$}
	\drawedge[ATnb=0,AHnb=1,ELside=r](fb7,fb){\small 0}
	\drawedge[ATnb=0,AHnb=1,ELside=l,linewidth=.5,AHLength=2,ELpos=70](fb2,fb3){\small $x_1[5]=1$}
	\drawedge[ATnb=0,AHnb=1,ELside=r](fb2,fa){\small 0}	
	
		\node[Nadjust=wh,Nmr=3](b0)(60,75){\small $b_7$} 
	\node[Nadjust=wh,Nmr=3](b1)(70,70){\small $b_6$} 
	\node[Nadjust=wh,Nmr=3](b2)(80,65){\small $b_5$} 
	\node[Nadjust=wh,Nmr=3](b3)(90,55){\small $b_4$} 
	\node[Nadjust=wh,Nmr=3](a3)(50,55){\small $b_4$} 
	\node[Nadjust=wh,Nmr=3](a4)(53,45){\small $b_3$} 
	\node[Nadjust=wh,Nmr=3](a5)(56,35){\small $b_2$} 
	\node[Nadjust=wh,Nmr=3](a6)(59,25){\small $b_1$} 
	\node[Nadjust=wh,Nmr=3](a7)(62,15){\small $b_0$} 
	\node[Nadjust=wh,Nmr=0](a8)(65,5){\small a} 
	\drawedge[ATnb=0,AHnb=1,ELside=r](b2,a3){\small 0} 
	\drawedge[ATnb=0,AHnb=1](a3,a4){\small 1} 
	\drawedge[ATnb=0,AHnb=1](a4,a5){\small 1} 	
	\drawedge[ATnb=0,AHnb=1](a5,a6){\small 1} 	
	\drawedge[ATnb=0,AHnb=1](a6,a7){\small 1} 	
	\drawedge[ATnb=0,AHnb=1](a7,a8){\small 1} 
	
	\node[Nadjust=wh,Nmr=3](b4)(97,45){\small $b_3$} 
	\node[Nadjust=wh,Nmr=3](b5)(100,35){\small $b_2$} 
	\node[Nadjust=wh,Nmr=3](b6)(103,25){\small $b_1$} 
	\node[Nadjust=wh,Nmr=3](b7)(106,15){\small $b_0$} 
	\node[Nadjust=wh,Nmr=0](b8)(103,5){\small d} 
	
	\node[Nadjust=wh,Nmr=3](c4)(87,45){\small $b_3$}

	\node[Nadjust=wh,Nmr=3,linewidth=.5](c5)(84,35){\small $b_2$} 
	\node[Nadjust=wh,Nmr=3](c6)(81,25){\small $b_1$}
	\node[Nadjust=wh,Nmr=3](c7)(78,15){\small $b_0$} 
	\node[Nadjust=wh,Nmr=0](c8)(75,5){\small b} 
	\node[Nadjust=wh,Nmr=0](c9)(81,5){\small c}

	\node[Nadjust=wh,Nmr=3](x6)(87,25){\small $b_1$} 
	\node[Nadjust=wh,Nmr=3](x7)(90,15){\small $b_0$} 
	\node[Nadjust=wh,Nmr=0](x8)(93,5){\small $x_1$} 	
         \drawedge[ATnb=0,AHnb=1](c5,x6){\small 1} 
          \drawedge[ATnb=0,AHnb=1](x6,x7){\small 1} 
           \drawedge[ATnb=0,AHnb=1](x7,x8){\small 1} 
	
	\drawedge[ATnb=0,AHnb=1](b0,b1){\small 1} 
	\drawedge[ATnb=0,AHnb=1](b1,b2){\small 1}  
	\drawedge[ATnb=0,AHnb=1](b2,b3){\small 1}
	\drawedge[ATnb=0,AHnb=1](b3,b4){\small 1}
	\drawedge[ATnb=0,AHnb=1](b4,b5){\small 1} 
	\drawedge[ATnb=0,AHnb=1](b5,b6){\small 1}   
	\drawedge[ATnb=0,AHnb=1](b6,b7){\small 1} 
	\drawedge[ATnb=0,AHnb=1](b7,b8){\small 0} 
	\drawedge[ATnb=0,AHnb=1,ELside=r](b3,c4){\small 0}
	\drawedge[ATnb=0,AHnb=1,ELside=r](c4,c5){\small 0} 
	\drawedge[ATnb=0,AHnb=1,ELside=r](c5,c6){\small 0}   
	\drawedge[ATnb=0,AHnb=1,ELside=r](c6,c7){\small 0} 
	\drawedge[ATnb=0,AHnb=1,ELside=r](c7,c8){\small 0} 
	\drawedge[ATnb=0,AHnb=1,ELside=l](c7,c9){\small 1}

	\node[Nframe=n,Nadjust=wh](x)(108,25){} 	
	\node[Nframe=n,Nadjust=wh](y)(118,25){} 	
	\drawedge[ATnb=0,AHnb=1,linewidth=.4,AHLength=2.5](x,y){}

	\node[Nframe=n,Nadjust=wh](z)(139,25){} 	
	\node[Nmr=0,Nadjust=wh](w)(149,18){$x_1$} 	
	\drawedge[ATnb=0,AHnb=1,linewidth=.4](z,w){1}

	\node[Nadjust=wh,Nmr=3](gb2)(135,45){\small $b_5$} 
	\node[Nadjust=wh,Nmr=3](gb3)(145,35){\small $b_4$} 
	\node[Nadjust=wh,Nmr=3](gb7)(135,15){\small $b_0$} 
	\node[Nadjust=wh,Nmr=0](gc)(155,25){\small $d$} 
	\node[Nadjust=wh,Nmr=0](gb)(132,5){\small $b$} 
	\node[Nadjust=wh,Nmr=0](gd)(140,5){\small $c$} 
	\node[Nframe=n,Nadjust=wh,Nmr=0](gx)(50,5){\small $x_1$} 
	\node[Nadjust=wh,Nmr=0](ga)(125,35){\small $a$} 
	\drawedge[ATnb=0,AHnb=1,ELside=r](gb3,gb7){\small 0}
	\drawedge[ATnb=0,AHnb=1,ELside=l](gb3,gc){\small 1}
	\drawedge[ATnb=0,AHnb=1,ELside=l](gb7,gd){\small 1}
	\drawedge[ATnb=0,AHnb=1,ELside=r](gb7,gb){\small 0}
	\drawedge[ATnb=0,AHnb=1,ELside=l](gb2,gb3){\small 1}
	\drawedge[ATnb=0,AHnb=1,ELside=r](gb2,ga){\small 0}	
	\put(115,55){$\rk{x_1}=\rk{c}+1$}

	\end{picture}		
	\setlength{\unitlength}{1mm}
	\caption{Obtenção do $\rk{x_1}$}
	\label{rk}
	\end{center}
\end{figure*}
Para encontrar o $\rk{x}$, é necessário realizar uma segunda busca. Do bit mais significativo
até o bit de interesse~$b'$, percorre-se a $\trc$ usando os bits de $x$. A partir do bit de interesse,
é necessário  encontrar o maior elemento da subárvore, ou seja, é necessário descer na árvore sempre para a direita, em direção ao maior elemento.

Para se obter este comportamento, será criada uma segunda chave de busca, $x'$ da seguinte maneira:

\begin{tabular}{rr}
& $x =x_{w-1}x_{w-2} \ldots x_2x_1x_0$\\
OR & $ 1~ 1 ~1 ~1 $  \\ \hline
&$x' = x_{w-1}x_{w-2} \ldots 1~1~1~1$
\end{tabular}

O número de 1's no final de $x'$ é   $b'$.  Ao trocar um bit de $x$ por $1$ a partir do bit de interesse, cria-se
uma uma chave de busca que irá encontrar o predecessor de $x$.
Seja $s''=\trs{x'}$. Então $\rk{x}=\rk{s''}+1$.
Veja um exemplo na Figura~\ref{rkd}. Observe que a máscara é  computável em $O(1)$ a partir de  $2^{b'+1}-1$.

\begin{figure*}[htb]
	\begin{center}
	\setlength{\unitlength}{1mm}
\begin{picture}(160,80)(0,0)
	\put(0,60){\begin{tabular}{cc}
	  S=\{a,b,c,d\}~~~~\\
	  a = 1 1 {\bf 0 1} 1 1 1 {\bf 1}\\
	  b = 1 1 {\bf 1 0}  0 0 0 {\bf 0}\\
	  c = 1 1 {\bf 1 0} 0 0 0 {\bf 1}\\
	  d = 1 1 {\bf 1 1} 1 1 1 {\bf 0}\\
	~\\		
	 $x_2$ = 1 1 {\bf 1 0} 1 0 0 {\bf 0}			
	\end{tabular}}

	\node[Nframe=n,Nadjust=wh](x)(33,25){} 	
	\node[Nframe=n,Nadjust=wh](y)(53,25){} 	
	\drawedge[ATnb=0,AHnb=1,linewidth=.4,AHLength=2.5](x,y){$\dlt{x_2}{b}=b_3$} 	

	\node[Nadjust=wh,Nmr=3](fb2)(15,35){\small $b_5$} 
	\node[Nadjust=wh,Nmr=3](fb3)(25,25){\small $b_4$} 
	\node[Nadjust=wh,Nmr=3](fb7)(15,15){\small $b_0$} 
	\node[Nadjust=wh,Nmr=0](fc)(35,15){\small $d$} 
	\node[Nadjust=wh,Nmr=0,linewidth=.4](fb)(5,5){\small $b$} 
	\node[Nadjust=wh,Nmr=0](fd)(25,5){\small $c$} 
	\node[Nframe=n,Nadjust=wh,Nmr=0](x)(0,5){\small  $x_2$} 
	\node[Nadjust=wh,Nmr=0](fa)(5,25){\small $a$} 
	\drawedge[ATnb=0,AHnb=1,ELside=r,linewidth=.4,AHLength=2.5,ELpos=70](fb3,fb7){\small $x_2[4]=0$}
	\drawedge[ATnb=0,AHnb=1,ELside=l](fb3,fc){\small 1}
	\drawedge[ATnb=0,AHnb=1,ELside=l](fb7,fd){\small $1$}
	\drawedge[ATnb=0,AHnb=1,ELside=r,linewidth=.4,AHLength=2.5,ELpos=70](fb7,fb){\small $x_2[0]=0$}
	\drawedge[ATnb=0,AHnb=1,ELside=l,linewidth=.4,AHLength=2.5,ELpos=70](fb2,fb3){\small $x_2[5]=1$}
	\drawedge[ATnb=0,AHnb=1,ELside=r](fb2,fa){\small 0}	
	
		\node[Nadjust=wh,Nmr=3](b0)(60,75){\small $b_7$} 
	\node[Nadjust=wh,Nmr=3](b1)(70,70){\small $b_6$} 
	\node[Nadjust=wh,Nmr=3](b2)(80,65){\small $b_5$} 
	\node[Nadjust=wh,Nmr=3](b3)(90,55){\small $b_4$} 
	\node[Nadjust=wh,Nmr=3](a3)(50,55){\small $b_4$} 
	\node[Nadjust=wh,Nmr=3](a4)(53,45){\small $b_3$} 
	\node[Nadjust=wh,Nmr=3](a5)(56,35){\small $b_2$} 
	\node[Nadjust=wh,Nmr=3](a6)(59,25){\small $b_1$} 
	\node[Nadjust=wh,Nmr=3](a7)(62,15){\small $b_0$} 
	\node[Nadjust=wh,Nmr=0](a8)(65,5){\small a} 
	\drawedge[ATnb=0,AHnb=1,ELside=r](b2,a3){\small 0} 
	\drawedge[ATnb=0,AHnb=1](a3,a4){\small 1} 
	\drawedge[ATnb=0,AHnb=1](a4,a5){\small 1} 	
	\drawedge[ATnb=0,AHnb=1](a5,a6){\small 1} 	
	\drawedge[ATnb=0,AHnb=1](a6,a7){\small 1} 	
	\drawedge[ATnb=0,AHnb=1](a7,a8){\small 1} 
	
	\node[Nadjust=wh,Nmr=3](b4)(97,45){\small $b_3$} 
	\node[Nadjust=wh,Nmr=3](b5)(100,35){\small $b_2$} 
	\node[Nadjust=wh,Nmr=3](b6)(103,25){\small $b_1$} 
	\node[Nadjust=wh,Nmr=3](b7)(106,15){\small $b_0$} 
	\node[Nadjust=wh,Nmr=0](b8)(103,5){\small d} 
	
	\node[Nadjust=wh,Nmr=3,linewidth=.5](c4)(87,45){\small $b_3$}

	\node[Nadjust=wh,Nmr=3](c5)(84,35){\small $b_2$} 
	\node[Nadjust=wh,Nmr=3](c6)(81,25){\small $b_1$}
	\node[Nadjust=wh,Nmr=3](c7)(78,15){\small $b_0$} 
	\node[Nadjust=wh,Nmr=0](c8)(75,5){\small b} 
	\node[Nadjust=wh,Nmr=0](c9)(81,5){\small c}

	\node[Nadjust=wh,Nmr=3](x5)(90,35){\small $b_2$} 
	\node[Nadjust=wh,Nmr=3](x6)(92,25){\small $b_1$} 
	\node[Nadjust=wh,Nmr=3](x7)(94,15){\small $b_0$} 
	\node[Nadjust=wh,Nmr=0](x8)(96,5){\small  $x_2$} 	
         \drawedge[ATnb=0,AHnb=1](x5,x6){\small 1} 
          \drawedge[ATnb=0,AHnb=1](x6,x7){\small 1} 
           \drawedge[ATnb=0,AHnb=1](x7,x8){\small 1} 
	
	\drawedge[ATnb=0,AHnb=1](c4,x5){\small 1} 
	
	\drawedge[ATnb=0,AHnb=1](b0,b1){\small 1} 
	\drawedge[ATnb=0,AHnb=1](b1,b2){\small 1}  
	\drawedge[ATnb=0,AHnb=1](b2,b3){\small 1}
	\drawedge[ATnb=0,AHnb=1](b3,b4){\small 1}
	\drawedge[ATnb=0,AHnb=1](b4,b5){\small 1} 
	\drawedge[ATnb=0,AHnb=1](b5,b6){\small 1}   
	\drawedge[ATnb=0,AHnb=1](b6,b7){\small 1} 
	\drawedge[ATnb=0,AHnb=1](b7,b8){\small 0} 
	\drawedge[ATnb=0,AHnb=1,ELside=r](b3,c4){\small 0}
	\drawedge[ATnb=0,AHnb=1,ELside=r](c4,c5){\small 0} 
	\drawedge[ATnb=0,AHnb=1,ELside=r](c5,c6){\small 0}   
	\drawedge[ATnb=0,AHnb=1,ELside=r](c6,c7){\small 0} 
	\drawedge[ATnb=0,AHnb=1,ELside=r](c7,c8){\small 0} 
	\drawedge[ATnb=0,AHnb=1,ELside=l](c7,c9){\small 1}

	\node[Nframe=n,Nadjust=wh](x)(108,25){} 	
	\node[Nframe=n,Nadjust=wh](y)(118,25){} 	
	\drawedge[ATnb=0,AHnb=1,linewidth=.4,AHLength=2.5](x,y){}


	\node[Nadjust=wh,Nmr=3](gb2)(135,45){\small $b_5$} 
	\node[Nadjust=wh,Nmr=3](gb3)(145,35){\small $b_4$} 
	\node[Nadjust=wh,Nmr=3](gb7)(135,23){\small $b_0$} 
	\node[Nadjust=wh,Nmr=0](gc)(155,25){\small $d$} 
	\node[Nadjust=wh,Nmr=0](gb)(132,10){\small $b$} 
	\node[Nadjust=wh,Nmr=0,linewidth=.5](gd)(140,10){\small $c$} 
	\node[Nframe=n,Nadjust=wh,Nmr=0](gx)(50,10){\small  x} 
	\node[Nadjust=wh,Nmr=0](ga)(125,35){\small $a$} 
	\drawedge[ATnb=0,AHnb=1,ELside=r,linewidth=.5,ELpos=70](gb3,gb7){\small $x_2'[4]=0$}
	\drawedge[ATnb=0,AHnb=1,ELside=l](gb3,gc){\small 1}
	\drawedge[ATnb=0,AHnb=1,ELside=l,linewidth=.5,ELpos=70](gb7,gd){\small $x_2'[0]=1$}
	\drawedge[ATnb=0,AHnb=1,ELside=r](gb7,gb){\small 0}
	\drawedge[ATnb=0,AHnb=1,ELside=l,linewidth=.5,ELpos=70](gb2,gb3){\small $x_2'[5]=1$}
	\drawedge[ATnb=0,AHnb=1,ELside=r](gb2,ga){\small 0}	
	\put(115,60){\begin{tabular}{cc}
	$x_2'$&$= x_2 ~OR~ 1111$ \\
	&$=1 1  {\bf 1 0}  ~ 1 1 1 {\bf 1}$
	\end{tabular}
	}
	\put(120,0){$\rk{x_2}=\rk{c}+1$}

	\end{picture}		
	\setlength{\unitlength}{1mm}
	\caption{Obtenção do $\rk{x_2}$ após a segunda busca na $\trc$}
	\label{rkd}
	\end{center}
\end{figure*}

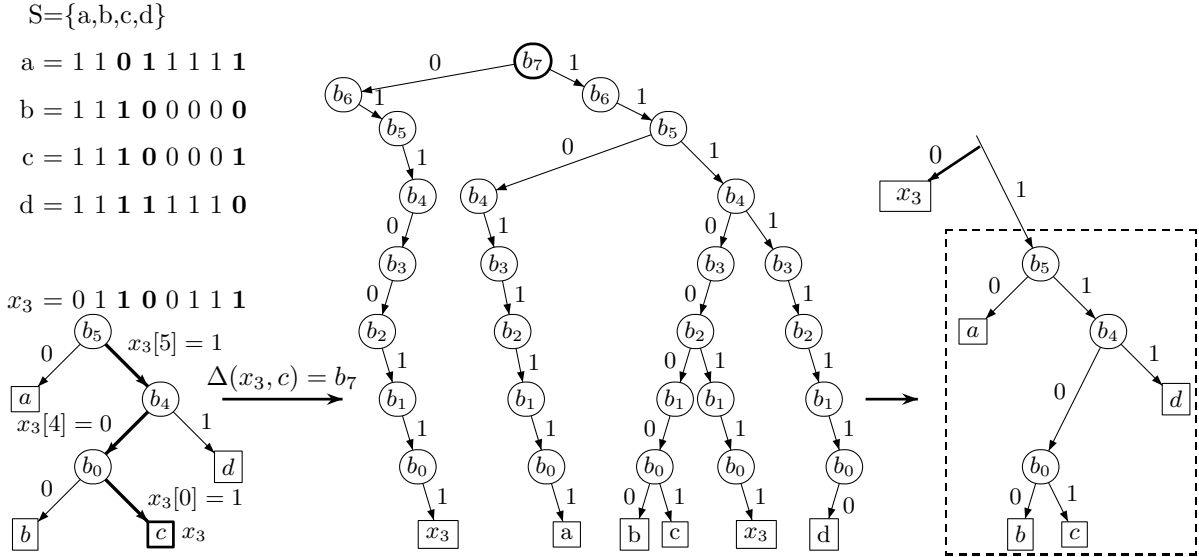
\begin{figure*}[htb]
	\begin{center}
	\setlength{\unitlength}{.9mm}
\begin{picture}(180,79)(0,0)
	\put(141,2){\dashbox(37,48)}	
	\put(0,60){\begin{tabular}{r}
	  S=\{a,b,c,d\}\hspace{1.07cm}~\\
	  a = 1 1 {\bf 0 1} 1 1 1 {\bf 1}\\
	  b = 1 1 {\bf 1 0}  0 0 0 {\bf 0}\\
	  c = 1 1 {\bf 1 0} 0 0 0 {\bf 1}\\
	  d = 1 1 {\bf 1 1} 1 1 1 {\bf 0}\\
	~\\		
	 $x_3$ = 0 1 {\bf 1 0} 0 1 1 {\bf 1}			
	\end{tabular}}

	\node[Nframe=n,Nadjust=wh](x)(33,25){} 	
	\node[Nframe=n,Nadjust=wh](y)(53,25){} 	
	\drawedge[ATnb=0,AHnb=1,linewidth=.4,AHLength=2.5](x,y){$\dlt{x_3}{c}=b_7$} 	

	\node[Nadjust=wh,Nmr=3](fb2)(15,35){\small $b_5$} 
	\node[Nadjust=wh,Nmr=3](fb3)(25,25){\small $b_4$} 
	\node[Nadjust=wh,Nmr=3](fb7)(15,15){\small $b_0$} 
	\node[Nadjust=wh,Nmr=0](fc)(35,15){\small $d$} 
	\node[Nadjust=wh,Nmr=0](fb)(5,5){\small $b$} 
	\node[Nadjust=wh,Nmr=0,linewidth=.4](fd)(25,5){\small $c$} 
	\node[Nframe=n,Nadjust=wh,Nmr=0](x)(30,5){\small  $x_3$} 
	\node[Nadjust=wh,Nmr=0](fa)(5,25){\small $a$} 
	\drawedge[ATnb=0,AHnb=1,ELside=r,linewidth=.5,AHLength=2,ELpos=90](fb3,fb7){\small $x_3[4]=0$}
	\drawedge[ATnb=0,AHnb=1,ELside=l](fb3,fc){\small 1}
	\drawedge[ATnb=0,AHnb=1,ELside=l,linewidth=.5,AHLength=2,ELpos=100](fb7,fd){\small $x_3[0]=1$}
	\drawedge[ATnb=0,AHnb=1,ELside=r](fb7,fb){\small 0}
	\drawedge[ATnb=0,AHnb=1,ELside=l,linewidth=.5,AHLength=2,ELpos=70](fb2,fb3){\small $x_3[5]=1$}
	\drawedge[ATnb=0,AHnb=1,ELside=r](fb2,fa){\small 0}

	\node[Nadjust=wh,Nmr=3](bb6)(52,70){\small $b_6$}
	\node[Nadjust=wh,Nmr=3](bb5)(60,65){\small $b_5$} 
	\node[Nadjust=wh,Nmr=3](bb4)(63,55){\small $b_4$} 
	\node[Nadjust=wh,Nmr=3](bb3)(60,45){\small $b_3$} 
	\node[Nadjust=wh,Nmr=3](bb2)(57,35){\small $b_2$} 
	\node[Nadjust=wh,Nmr=3](bb1)(60,25){\small $b_1$} 
	\node[Nadjust=wh,Nmr=3](bb0)(63,15){\small $b_0$} 
	\node[Nadjust=wh,Nmr=0](x3)(66,5){\small  $x_3$}	
	\drawedge[ATnb=0,AHnb=1](bb6,bb5){\small 1} 
	\drawedge[ATnb=0,AHnb=1](bb5,bb4){\small 1} 
	\drawedge[ATnb=0,AHnb=1,ELside=r](bb4,bb3){\small 0} 
	\drawedge[ATnb=0,AHnb=1,ELside=r](bb3,bb2){\small 0} 	
	\drawedge[ATnb=0,AHnb=1](bb2,bb1){\small 1} 	
	\drawedge[ATnb=0,AHnb=1](bb1,bb0){\small 1} 	
	\drawedge[ATnb=0,AHnb=1](bb0,x3){\small 1} 		
		\node[Nadjust=wh,Nmr=3,linewidth=.4](b0)(80,75){\small $b_7$} 
		\drawedge[ATnb=0,AHnb=1,ELside=r](b0,bb6){\small 0} 
		
	\node[Nadjust=wh,Nmr=3](b1)(90,70){\small $b_6$} 
	\node[Nadjust=wh,Nmr=3](b2)(100,65){\small $b_5$} 
	\node[Nadjust=wh,Nmr=3](b3)(110,55){\small $b_4$} 
	\node[Nadjust=wh,Nmr=3](a3)(72,55){\small $b_4$} 
	\node[Nadjust=wh,Nmr=3](a4)(75,45){\small $b_3$} 
	\node[Nadjust=wh,Nmr=3](a5)(77,35){\small $b_2$} 
	\node[Nadjust=wh,Nmr=3](a6)(79,25){\small $b_1$} 
	\node[Nadjust=wh,Nmr=3](a7)(82,15){\small $b_0$} 
	\node[Nadjust=wh,Nmr=0](a8)(85,5){\small a} 
	\drawedge[ATnb=0,AHnb=1,ELside=r](b2,a3){\small 0} 
	\drawedge[ATnb=0,AHnb=1](a3,a4){\small 1} 
	\drawedge[ATnb=0,AHnb=1](a4,a5){\small 1} 	
	\drawedge[ATnb=0,AHnb=1](a5,a6){\small 1} 	
	\drawedge[ATnb=0,AHnb=1](a6,a7){\small 1} 	
	\drawedge[ATnb=0,AHnb=1](a7,a8){\small 1} 
	
	\node[Nadjust=wh,Nmr=3](b4)(117,45){\small $b_3$} 
	\node[Nadjust=wh,Nmr=3](b5)(120,35){\small $b_2$} 
	\node[Nadjust=wh,Nmr=3](b6)(123,25){\small $b_1$} 
	\node[Nadjust=wh,Nmr=3](b7)(126,15){\small $b_0$} 
	\node[Nadjust=wh,Nmr=0](b8)(123,5){\small d} 
	
	\node[Nadjust=wh,Nmr=3](c4)(107,45){\small $b_3$}

	\node[Nadjust=wh,Nmr=3](c5)(104,35){\small $b_2$} 
	\node[Nadjust=wh,Nmr=3](c6)(101,25){\small $b_1$}
	\node[Nadjust=wh,Nmr=3](c7)(98,15){\small $b_0$} 
	\node[Nadjust=wh,Nmr=0](c8)(95,5){\small b} 
	\node[Nadjust=wh,Nmr=0](c9)(101,5){\small c}

	\node[Nadjust=wh,Nmr=3](x6)(107,25){\small $b_1$} 
	\node[Nadjust=wh,Nmr=3](x7)(110,15){\small $b_0$} 
	\node[Nadjust=wh,Nmr=0](x8)(113,5){\small  $x_3$} 	
         \drawedge[ATnb=0,AHnb=1](c5,x6){\small 1} 
          \drawedge[ATnb=0,AHnb=1](x6,x7){\small 1} 
           \drawedge[ATnb=0,AHnb=1](x7,x8){\small 1} 
	
	\drawedge[ATnb=0,AHnb=1](b0,b1){\small 1} 
	\drawedge[ATnb=0,AHnb=1](b1,b2){\small 1}  
	\drawedge[ATnb=0,AHnb=1](b2,b3){\small 1}
	\drawedge[ATnb=0,AHnb=1](b3,b4){\small 1}
	\drawedge[ATnb=0,AHnb=1](b4,b5){\small 1} 
	\drawedge[ATnb=0,AHnb=1](b5,b6){\small 1}   
	\drawedge[ATnb=0,AHnb=1](b6,b7){\small 1} 
	\drawedge[ATnb=0,AHnb=1](b7,b8){\small 0} 
	\drawedge[ATnb=0,AHnb=1,ELside=r](b3,c4){\small 0}
	\drawedge[ATnb=0,AHnb=1,ELside=r](c4,c5){\small 0} 
	\drawedge[ATnb=0,AHnb=1,ELside=r](c5,c6){\small 0}   
	\drawedge[ATnb=0,AHnb=1,ELside=r](c6,c7){\small 0} 
	\drawedge[ATnb=0,AHnb=1,ELside=r](c7,c8){\small 0} 
	\drawedge[ATnb=0,AHnb=1,ELside=l](c7,c9){\small 1}

	\node[Nframe=n,Nadjust=wh](x)(128,25){} 	
	\node[Nframe=n,Nadjust=wh](y)(138,25){} 	
	\drawedge[ATnb=0,AHnb=1,linewidth=.4,AHLength=2.5](x,y){}

	\node[Nadjust=wh,Nmr=3](gb2)(155,45){\small $b_5$} 
	\node[Nadjust=wh,Nmr=3](gb3)(165,35){\small $b_4$} 
	\node[Nadjust=wh,Nmr=3](gb7)(155,15){\small $b_0$} 
	\node[Nadjust=wh,Nmr=0](gc)(175,25){\small $d$} 
	\node[Nadjust=wh,Nmr=0](gb)(152,5){\small $b$} 
	\node[Nadjust=wh,Nmr=0](gd)(160,5){\small $c$} 
	\node[Nadjust=wh,Nmr=0](ga)(145,35){\small $a$} 
	\drawedge[ATnb=0,AHnb=1,ELside=r](gb3,gb7){\small 0}
	\drawedge[ATnb=0,AHnb=1,ELside=l](gb3,gc){\small 1}
	\drawedge[ATnb=0,AHnb=1,ELside=l](gb7,gd){\small 1}
	\drawedge[ATnb=0,AHnb=1,ELside=r](gb7,gb){\small 0}
	\drawedge[ATnb=0,AHnb=1,ELside=l](gb2,gb3){\small 1}
	\drawedge[ATnb=0,AHnb=1,ELside=r](gb2,ga){\small 0}	

	\node[Nframe=n,Nadjust=wh](z2)(145,65){} 	
	\node[Nframe=n,Nadjust=wh](z)(147,63){} 	
	\node[Nmr=0,Nadjust=wh](w)(135,55){ $x_3$} 	
	\drawedge[ATnb=0,AHnb=1,linewidth=.4,ELside=r](z,w){0} 	
	\drawedge[ATnb=0,AHnb=1,ELside=l](z2,gb2){\small 1}

	\end{picture}		
	\setlength{\unitlength}{1mm}
	\caption{Obtenção do $\rk{x_3}$. O menor elemento da sub-árvore em destaque é o sucessor de $x_3$.}
	\label{rkb}
	\end{center}
\end{figure*}

{\bf Caso  b ($x[b']=0$)}
No exemplo da Figura~\ref{rkb}, verifica-se que o sucessor de $x_3$ é o menor elemento da 
sub-árvore em destaque.
Para encontrar o $\rk{x_3}$, é necessário realizar uma segunda busca. Do bit mais significativo
até o bit de interesse~$b'$, percorre-se a $\trc$ usando os bits de $x_3$. A partir do bit de interesse,
continua-se descendo na árvore sempre para a esquerda, em direção ao menor elemento. 

Para obter este comportamento, será criada uma segunda chave de busca, $x'$ da seguinte maneira:

\begin{tabular}{rr}
& $x =x_{w-1}x_{w-2} \ldots x_2x_1x_0$\\
AND & $ 1~ 1 ~1 ~1\ldots0~ 0 ~0 ~0 $  \\ \hline
&$x' = x_{w-1}x_{w-2} \ldots 0~0~0~0$
\end{tabular}

O número de 0's no final de $x'$ é   $b'$.  Ao trocar um bit de $x$ por $0$ a partir do bit de interesse, cria-se
uma uma chave de busca que irá encontrar o predecessor de $x$.
Seja $s''=\trs{x'}$. Então $\rk{x}=\rk{s''}$.
Veja um exemplo na Figura~\ref{rkt}.

\begin{figure*}[htb]
	\begin{center}
	\setlength{\unitlength}{1mm}
\begin{picture}(160,80)(0,0)
	\put(0,60){\begin{tabular}{rl}
	 & S=\{a,b,c,d\}\\
	  a\hspace{-.3cm}&\hspace{-.3cm}=1 1 {\bf 0 1} 1 1 1 {\bf 1}\\
	  b\hspace{-.3cm}&\hspace{-.3cm}=1 1 {\bf 1 0}  0 1 0 {\bf 0}\\
	  c\hspace{-.3cm}&\hspace{-.3cm}=1 1 {\bf 1 0} 0 1 0 {\bf 1}\\
	  d\hspace{-.3cm}&\hspace{-.3cm}=1 1 {\bf 1 1} 1 1 1 {\bf 0}\\
			
	 $x_4$\hspace{-.3cm}&\hspace{-.3cm}=1 1 {\bf 1 0} 0 0 0 {\bf 1}			
	\end{tabular}}

	\node[Nframe=n,Nadjust=wh](x)(33,25){} 	
	\node[Nframe=n,Nadjust=wh](y)(53,25){} 	
	\drawedge[ATnb=0,AHnb=1,linewidth=.4,AHLength=2.5](x,y){$\dlt{x_4}{c}=b_2$} 	

	\node[Nadjust=wh,Nmr=3](fb2)(15,35){\small $b_5$} 
	\node[Nadjust=wh,Nmr=3](fb3)(25,25){\small $b_4$} 
	\node[Nadjust=wh,Nmr=3](fb7)(15,15){\small $b_0$} 
	\node[Nadjust=wh,Nmr=0](fc)(35,15){\small $d$} 
	\node[Nadjust=wh,Nmr=0](fb)(5,5){\small $b$} 
	\node[Nadjust=wh,Nmr=0,linewidth=.4](fd)(25,5){\small $c$} 
	\node[Nframe=n,Nadjust=wh,Nmr=0](x)(0,5){\small  $x_4$} 
	\node[Nadjust=wh,Nmr=0](fa)(5,25){\small $a$} 
	\drawedge[ATnb=0,AHnb=1,ELside=r,linewidth=.4,AHLength=2.5,ELpos=90](fb3,fb7){\small $x_4[4]=0$}
	\drawedge[ATnb=0,AHnb=1,ELside=l](fb3,fc){\small 1}
	\drawedge[ATnb=0,AHnb=1,ELside=l,linewidth=.4,AHLength=2.5,ELpos=90](fb7,fd){\small $x_4[0]=1$}
	\drawedge[ATnb=0,AHnb=1,ELside=r](fb7,fb){\small $0$}
	\drawedge[ATnb=0,AHnb=1,ELside=l,linewidth=.4,AHLength=2.5,ELpos=70](fb2,fb3){\small $x_4[5]=1$}
	\drawedge[ATnb=0,AHnb=1,ELside=r](fb2,fa){\small 0}	
	
		\node[Nadjust=wh,Nmr=3](b0)(60,75){\small $b_7$} 
	\node[Nadjust=wh,Nmr=3](b1)(70,70){\small $b_6$} 
	\node[Nadjust=wh,Nmr=3](b2)(80,65){\small $b_5$} 
	\node[Nadjust=wh,Nmr=3](b3)(90,55){\small $b_4$} 
	\node[Nadjust=wh,Nmr=3](a3)(50,55){\small $b_4$} 
	\node[Nadjust=wh,Nmr=3](a4)(53,45){\small $b_3$} 
	\node[Nadjust=wh,Nmr=3](a5)(56,35){\small $b_2$} 
	\node[Nadjust=wh,Nmr=3](a6)(59,25){\small $b_1$} 
	\node[Nadjust=wh,Nmr=3](a7)(62,15){\small $b_0$} 
	\node[Nadjust=wh,Nmr=0](a8)(65,5){\small a} 
	\drawedge[ATnb=0,AHnb=1,ELside=r](b2,a3){\small 0} 
	\drawedge[ATnb=0,AHnb=1](a3,a4){\small 1} 
	\drawedge[ATnb=0,AHnb=1](a4,a5){\small 1} 	
	\drawedge[ATnb=0,AHnb=1](a5,a6){\small 1} 	
	\drawedge[ATnb=0,AHnb=1](a6,a7){\small 1} 	
	\drawedge[ATnb=0,AHnb=1](a7,a8){\small 1} 
	
	\node[Nadjust=wh,Nmr=3](b4)(97,45){\small $b_3$} 
	\node[Nadjust=wh,Nmr=3](b5)(100,35){\small $b_2$} 
	\node[Nadjust=wh,Nmr=3](b6)(103,25){\small $b_1$} 
	\node[Nadjust=wh,Nmr=3](b7)(106,15){\small $b_0$} 
	\node[Nadjust=wh,Nmr=0](b8)(103,5){\small d} 
	
	\node[Nadjust=wh,Nmr=3](c4)(87,45){\small $b_3$}

	\node[Nadjust=wh,Nmr=3,linewidth=.5](c5)(84,35){\small $b_2$} 
	\node[Nadjust=wh,Nmr=3](c6)(93,25){\small $b_1$}
	\node[Nadjust=wh,Nmr=3](c7)(87,15){\small $b_0$} 
	\node[Nadjust=wh,Nmr=0](c8)(84,5){\small b} 
	\node[Nadjust=wh,Nmr=0](c9)(90,5){\small c}

	\node[Nadjust=wh,Nmr=3](x6)(79,25){\small $b_1$} 
	\node[Nadjust=wh,Nmr=3](x7)(75,15){\small $b_0$} 
	\node[Nadjust=wh,Nmr=0](x8)(75,5){\small  $x_4$} 	
         \drawedge[ATnb=0,AHnb=1,ELside=r](c5,x6){\small 0} 
          \drawedge[ATnb=0,AHnb=1,ELside=r](x6,x7){\small 0} 
           \drawedge[ATnb=0,AHnb=1](x7,x8){\small 1} 
%
	
	\drawedge[ATnb=0,AHnb=1](b0,b1){\small 1} 
	\drawedge[ATnb=0,AHnb=1](b1,b2){\small 1}  
	\drawedge[ATnb=0,AHnb=1](b2,b3){\small 1}
	\drawedge[ATnb=0,AHnb=1](b3,b4){\small 1}
	\drawedge[ATnb=0,AHnb=1](b4,b5){\small 1} 
	\drawedge[ATnb=0,AHnb=1](b5,b6){\small 1}   
	\drawedge[ATnb=0,AHnb=1](b6,b7){\small 1} 
	\drawedge[ATnb=0,AHnb=1](b7,b8){\small 0} 
	\drawedge[ATnb=0,AHnb=1,ELside=r](b3,c4){\small 0}
	\drawedge[ATnb=0,AHnb=1,ELside=r](c4,c5){\small 0} 
	\drawedge[ATnb=0,AHnb=1,ELside=l](c5,c6){\small 1}   
	\drawedge[ATnb=0,AHnb=1,ELside=r](c6,c7){\small 0} 
	\drawedge[ATnb=0,AHnb=1,ELside=r](c7,c8){\small 0} 
	\drawedge[ATnb=0,AHnb=1,ELside=l](c7,c9){\small 1}

	\node[Nframe=n,Nadjust=wh](x)(108,25){} 	
	\node[Nframe=n,Nadjust=wh](y)(118,25){} 	
	\drawedge[ATnb=0,AHnb=1,linewidth=.4,AHLength=2.5](x,y){}


	\node[Nadjust=wh,Nmr=3](gb2)(135,45){\small $b_5$} 
	\node[Nadjust=wh,Nmr=3](gb3)(145,35){\small $b_4$} 
	\node[Nadjust=wh,Nmr=3](gb7)(135,23){\small $b_0$} 
	\node[Nadjust=wh,Nmr=0](gc)(155,25){\small $d$} 
	\node[Nadjust=wh,Nmr=0,,linewidth=.5](gb)(132,10){\small $b$} 
	\node[Nadjust=wh,Nmr=0](gd)(140,10){\small $c$} 
	\node[Nframe=n,Nadjust=wh,Nmr=0](gx)(50,10){\small  x} 
	\node[Nadjust=wh,Nmr=0](ga)(125,35){\small $a$} 
	\drawedge[ATnb=0,AHnb=1,ELside=r,linewidth=.5,ELpos=70](gb3,gb7){\small $x_4'[4]=0$}
	\drawedge[ATnb=0,AHnb=1,ELside=l](gb3,gc){\small 1}
	\drawedge[ATnb=0,AHnb=1,ELside=l](gb7,gd){\small $1$}
	\drawedge[ATnb=0,AHnb=1,ELside=r,linewidth=.5,ELpos=70](gb7,gb){\small $x_4'[0]=0$}
	\drawedge[ATnb=0,AHnb=1,ELside=l,linewidth=.5,ELpos=70](gb2,gb3){\small $x_4'[5]=1$}
	\drawedge[ATnb=0,AHnb=1,ELside=r](gb2,ga){\small 0}	
	\put(115,60){\begin{tabular}{rl}
	$x_4'$&$= x_4 ~AND~ 1 1 1 1 1 0 0 0 $ \\
	&$=1 1 {\bf 1 0} 0 0 0 {\bf 0}$
	\end{tabular}
	}
	\put(120,0){$\rk{x_4}=\rk{b}$}

	\end{picture}		
	\setlength{\unitlength}{1mm}
	\caption{Obtenção do $\rk{x_4}$ após a segunda busca na $\trc$}
	\label{rkt}
	\end{center}
\end{figure*}

\subsection{Característica da Árvore de Fusão}

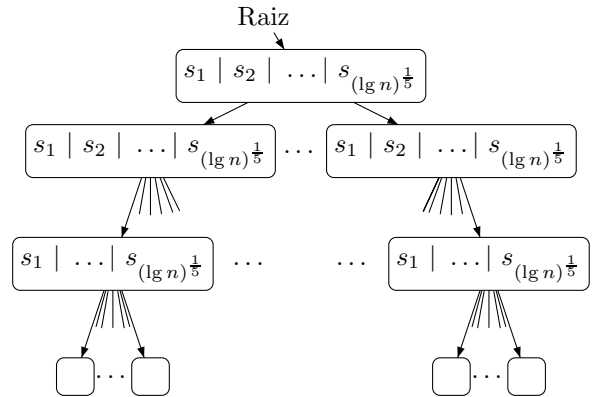
\begin{figure}[htb!]
	\begin{center}
		\begin{picture}(80,50)(0,0)

	\node[Nframe=n,Nadjust=wh,Nmr=1](R)(35,48){Raiz} 
	\node[Nadjust=wh,Nmr=1](n1)(40,40){$s_1~|~s_2~|~\ldots|~s_{(\lg n)^\frac{1}{5}}$} 
	\drawedge[ATnb=0,AHnb=1](R,n1){} 
	\node[Nadjust=wh,Nmr=1](n2)(20,30){$s_1~|~s_2~|~\ldots|~s_{(\lg n)^\frac{1}{5}}$} 
	\node[Nadjust=wh,Nmr=1](n3)(60,30){$s_1~|~s_2~|~\ldots|~s_{(\lg n)^\frac{1}{5}}$} 
	\node[Nframe=n,Nadjust=wh,Nmr=1](x)(40,30){\ldots} 
	\drawedge[ATnb=0,AHnb=1](n1,n2){} 
	\drawedge[ATnb=0,AHnb=1](n1,n3){} 
	\node[Nadjust=wh,Nmr=1](n4)(15,15){$s_1~|~\ldots|~s_{(\lg n)^\frac{1}{5}}$} 
	\node[Nframe=n,Nadjust=wh,Nmr=1](n6)(33,15){$\ldots$} 
	\drawedge[ATnb=0,AHnb=1](n2,n4){}

	\imark[iangle=260,ATnb=0,AHnb=0](n2)
	\imark[iangle=270,ATnb=0,AHnb=0](n2)
	\imark[iangle=280,ATnb=0,AHnb=0](n2)
	\imark[iangle=290,ATnb=0,AHnb=0](n2)
	\imark[iangle=296,ATnb=0,AHnb=0](n2)

	\imark[iangle=245,ATnb=0,AHnb=0](n3)
	\imark[iangle=250,ATnb=0,AHnb=0](n3)
	\imark[iangle=260,ATnb=0,AHnb=0](n3)
	\imark[iangle=270,ATnb=0,AHnb=0](n3)
	\imark[iangle=280,ATnb=0,AHnb=0](n3)
	
	\node[Nframe=n,Nadjust=wh,Nmr=1](n7)(47,15){\ldots} 

	\node[Nadjust=wh,Nmr=1](n9)(65,15){$s_1~|~\ldots|~s_{(\lg n)^\frac{1}{5}}$} 
	\drawedge[ATnb=0,AHnb=1](n3,n9){}  
		\imark[iangle=245,ATnb=0,AHnb=0](n3)
	
		\imark[iangle=245,ATnb=0,AHnb=0](n3)

	\imark[iangle=255,ATnb=0,AHnb=0](n4)
	\imark[iangle=260,ATnb=0,AHnb=0](n4)
	\imark[iangle=270,ATnb=0,AHnb=0](n4)
	\imark[iangle=280,ATnb=0,AHnb=0](n4)
	\imark[iangle=285,ATnb=0,AHnb=0](n4)

	\imark[iangle=255,ATnb=0,AHnb=0](n9)
	\imark[iangle=260,ATnb=0,AHnb=0](n9)
	\imark[iangle=270,ATnb=0,AHnb=0](n9)
	\imark[iangle=280,ATnb=0,AHnb=0](n9)
	\imark[iangle=285,ATnb=0,AHnb=0](n9)

	\node[Nw=5,Nh=5,Nmr=1](n10)(10,0){} 
	\node[Nframe=n,Nw=5,Nh=5,Nmr=1](x)(15,0){\ldots} 
	\node[Nw=5,Nh=5,Nmr=1](n11)(20,0){}
	\drawedge[ATnb=0,AHnb=1](n4,n10){}
	\drawedge[ATnb=0,AHnb=1](n4,n11){} 

	\node[Nw=5,Nh=5,Nmr=1](n16)(60,0){} 
	\node[Nframe=n,Nw=5,Nh=5,Nmr=1](x)(65,0){\ldots} 
	\node[Nw=5,Nh=5,Nmr=1](n17)(70,0){} 

	\drawedge[ATnb=0,AHnb=1](n9,n16){}
	\drawedge[ATnb=0,AHnb=1](n9,n17){} 

	\end{picture}	

	\caption{Estrutura de uma Árvore de Fusão Completa.}
	\label{fig_arvore_fusao}
	\end{center}
\end{figure}

Basicamente, uma Árvore de Fusão é uma árvore B com grau $B = (\lg n)^\frac{1}{5}$, ou seja, o grau é crescente com relação à quantidade de números de itens a serem ordenados, como exemplificado na Figura~\ref{fig_arvore_fusao}. 

Como a altura de uma árvore B é proporcional a   $\log_B n $ e na árvore de fusão $B = (\lg n)^\frac{1}{5}$ então a altura h
é O de:

\[ \log_B n = \frac{\lg n}{\lg B} = \frac{\lg n}{\lg (\lg n)^{1/5}} = \cdot \frac{\lg n}{\lg \frac{1}{5}\lg n} = O(\frac{\lg n}{\lg \lg n}) \]


O tempo de busca dentro de um nó de uma árvore B é $O( B)$, pois é feita uma busca sequencial para encontrar o filho. Isso em cada nível da árvore, resultando em $O(B \log_B n)$.
Numa Árvore de Fusão, o nó filho é encontrado em $O(1)$. Assim, a busca termina em $O(\log_B n)$. Como $B =(\lg n)^\frac{1}{5}$, a busca ocorre em

\[  \log_B n = O(\dfrac{\lg n}{\lg \lg n})\]




Como visto anteriormente, para ordenar números binários, não é necessário comparar todos os bits, mas somente os bits de interesse. 
Esses bits de interesse  são chamados de sketch:

\begin{definicao}{\sk{s}:}
O \textit{sketch} de uma palavra $s$ consiste em descartar todos os seus bits, exceto os bits de interesse. A ordem das palavras são preservadas, ou seja, $s_i < s_j$ se e somente se $\sk{s_i} < \sk{s_j}$. 
\end{definicao}

Por exemplo, na Figura \ref{fig_pesquisaTrie}, os \textit{sketches} dos elementos $a$, $b$, $c$ e $d$ serão respectivamente $011$, $100$, $101$ e $110$. E pode-se perceber que a ordem entre os \textit{sketches} não muda em relação aos números originais.

A ideia central da Árvore de Fusão está em como ela armazena as chaves em cada nó. 
Cada um de seus nós possui $t\leq B-1 = O(w^{\frac{1}{5}})$ chaves. Conforme o Lema~\ref{nbits}, em um conjunto de $B-1$ chaves, a 
$\tr$ terá, no máximo, $B-2$ bits de interesse. Um nó poussui no máximo $B-1$ \textit{sketches}, cada \textit{sketch}  possui no máximo $B-2$ bits de interesse. Então, a soma dos bits dos \textit{sketches} será:

$$ (B-1)\cdot (B-2) \leq w^\frac{1}{5} \cdot w^\frac{1}{5} = O(w^\frac{2}{5}) = o(w) = o(\lg n).$$

Concluímos, então, que a soma dos bits dos \textit{sketches} das chaves em um nó cabem em apenas uma palavra da memória. 
Cada nó da árvore conterá uma palavra  que armazenará todos os \textit{sketches} das chaves e mais alguns bits, 
conforme a definição a seguir:

\begin{definicao}{Nó Sketch:}
Nó \textit{Sketch} é um nó que contém todos os \textit{sketches} das chaves $(s_1,\ldots,s_t)$. Estes \textit{sketches} podem ser armazenados em uma única palavra, sendo que cada \textit{sketch} é precedido de um bit
separador com valor~1. O Nó Sketch será a concatenação dos sketchs das chaves: $$w_{node} = 1\sk{s_1}1\sk{s_2} ... 1\sk{s_t}.$$ 
Além disso, \textit{sketches} são concatenados em ordem crescente.
\end{definicao}

Na próxima seção será mostrado a comparação de um dado número $x$ com todas as chaves de um nó em tempo constante, baseado em~\cite{lec2011}.

\subsection{Múltiplas Comparações em tempo constante}

%


Considere um nó da árvore de fusão com elementos $S=(s_1,\ldots,s_t)$.  Suponha que os bits 
de interesse deste conjunto sejam $(i_1,\ldots,i_{t'})$ com $t'<t$.
Para comparar uma chave de pesquisa $x$ com todas as chaves de um nó, primeiramente computa-se \textit{\sk{x}}.

Para extrair o bit de interesse $i_1$ e colocá-lo na posição 0 de um  $\sk{x}$, primeiro é feito um $AND$ bit a bit 
de $x$ e uma máscara contendo zeros e um único 1 na posição $i_1$ . Uma vez aplicada esta máscara, é necessário mover o bit para a posição 0. Isso pode ser feito 
por um deslocamento $delta$ para a esquerda, obtida por uma multiplicação por $2^{delta}$.

Para se obter todos os bits do $\sk{x}$ é necessário fazer o $AND$ bit a bit de $x$ 
com uma máscara contendo valores 1 apenas nos bits de interesse  $(i_1,\ldots,i_{t'})$.
Esta máscara  será construída junto com a $\trc$ e estará disponível no momento
da busca de $x$.
\newpage

Após aplicar a máscara, deve-se mover os bits para as posições iniciais criando o  $\sk{x}$ em $O(1)$ conforme a figura abaixo:

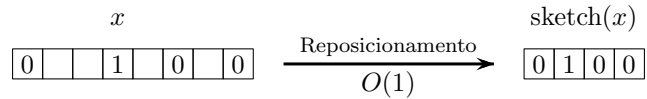
\centerline{
\begin{picture}(110,20)(0,0)
\node[Nmr=0,Nw=4,Nh=4,linecolor=white](B)(24,16){$x$}
\node[Nmr=0,Nw=4,Nh=4,linecolor=white](B)(86,16){$\sk{x}$}
\node[Nmr=0,Nw=4,Nh=4](B)(12,10){0}
\node[Nmr=0,Nw=4,Nh=4](B)(16,10){}
\node[Nmr=0,Nw=4,Nh=4](B)(20,10){}
\node[Nmr=0,Nw=4,Nh=4](B)(24,10){1}
\node[Nmr=0,Nw=4,Nh=4](B)(28,10){}
\node[Nmr=0,Nw=4,Nh=4](B)(32,10){0}
\node[Nmr=0,Nw=4,Nh=4](B)(36,10){}
\node[Nmr=0,Nw=4,Nh=4](B)(40,10){0}
\node[Nframe=n,Nadjust=wh](Y)(75,10){} 	
\node[Nframe=n,Nadjust=wh](X)(45,10){} 	
\drawedge[ATnb=0,AHnb=1,linewidth=.4,AHLength=2.5](X,Y){\footnotesize Reposicionamento}
\drawedge[ELside=r](X,Y){$O(1)$} 			 	
\node[Nmr=0,Nw=4,Nh=4](B)(80,10){0}
\node[Nmr=0,Nw=4,Nh=4](B)(84,10){1}
\node[Nmr=0,Nw=4,Nh=4](B)(88,10){0}
\node[Nmr=0,Nw=4,Nh=4](B)(92,10){0}
\end{picture}
}

 Ao multiplicar um inteiro $x$ por uma constante pré-definida, é possível alterar
a posição dos bits de $x$. Obter este reposicionamento de bits com uma única multiplicação não é tarefa trivial. 
O trabalho \cite{lec2012} mostra a existência de constantes pré-definidas que obtêm um reposicionamento 
 dos bits. O resultado não é perfeito, pois as constantes geram alguns zeros adicionais no sketch. 
 Tais zeros são criados de forma a não alterar o funciomento do algoritmo.

Uma vez obtido o $\sk{x}$, seu 
valor será replicado dentro de uma palavra acrescido do bit separador 0 da seguinte maneira:
 $$w_x = 0~\sk{x}~0~\sk{x} ~...~ 0~\sk{x}$$

Suponha que $\sk{x}$ possui 6 bits, assim tem-se que

\begin{equation*} 
\begin{aligned}
w_x &= \sk{x} + \sk{x}\cdot 2^7 + \sk{x} \cdot 2^{14} + ~... \\
& = \sk{x}\cdot(...10000010000001).\\
\end{aligned}
\end{equation*} 

Ou seja, é possível calcular $w_x$ a partir de $\sk{x}$ com uma única multiplicação.

\begin{fato}
Ao sutrair   $1\sk{s_i}-0\sk{x}$, o resultado começará com 1 se e somente se $\sk{x}\leq \sk{s_i}$. 
\end{fato}

Seja $\sk{x} = 1111$ e $\sk{s_i} = 0000$, assim,
$1\sk{s_i} - 0\sk{x} ~ = ~ 10000 - 01111 = \mathbf{0}0001$.
Como a subtração começa com zero, então $\sk{x}>\sk{s_i}  $.

Agora, se $\sk{x}=0000$ e $\sk{s_i}=00001$, tem-se
$1\sk{s_i} - 0\sk{x} = 10001-00000 = \mathbf{1}0001$.
Como o primeiro bit é 1, $ \sk{x}\leq \sk{s_i}  $.

Para comparar $x$ com todas as chaves de S em $O(1)$,  é necessário calcular a subtração entre $1\sk{s} - 0\sk{x}$ 
para todos $s\in S$ em uma única operação. Ou seja, múltiplas comparações em uma única operação. Calcula-se   $$w_{res} = w_{node}-w_x.$$

O primeiro bit de cada bloco indicará se $\sk{x}$ é menor igual ou maior que $\sk{s_i}$. Vale a pena lembrar que os $sketches$ estão ordenados dentro do \textit{nó sketch} $w_{node}$.
É necessário encontrar  o primeiro bit de cada bloco que valha 1.
Suponha que o tamanho do bloco $0\sk{x}$  é $r$.
 Para remover todos os bits, exceto os primeiros bits de cada bloco, será feita uma operação \textit{AND bit-a-bit} entre $w_{res}$ e uma máscara contendo 1 somente nas posições de interesse, $(r,2r,\ldots)$.  Seja $w'_{res}$  o resultado deste
 AND bit a bit.

O próximo passo  é encontrar o bit mais significativo igual a 1. Esta operação equivale ao calculo de 
$\lfloor\lg (w'_{res})\rfloor$ e precisa ser realizada em $O(1)$.
A solução deste problema é conhecida na literatura~\cite{hackers}.

O elemento $s=\trs{x}$ pode ser computado diretamente a partir da posição do primeiro 1 no início de um bloco em $w_{res}$. 
Seguindo os passos da sessão anterior  tem-se o $\rk{x}$ em $O(1)$. Usando esta operação, descobre-se qual filho seguir na busca dentro da Árvore de Fusão em $O(1)$.

Como exemplo, será calculado $\trs{x}$, com os mesmos valores da Figura~\ref{fig_pesquisaTrie}. 
$$S=(a,b,c,d)=(11011111, 11100000, 11100001, 11111110)=(223,224,225,245)$$

Considere que $x = 11100111=231$ para esse caso. Os \textit{sketches} dos elementos serão

$$ \sk{a} = 011;~~~ \sk{b} = 100; $$

$$\sk{c} = 101;~~~ \sk{d} = 110  $$

o nó \textit{sketch} será

$$ w_{node} = 1~011~1~100~1~101~1~110=48350 $$

e o $\sk{x}$ será $101$. A palavra para subtrair de $ w_{node}$ será:

$$ w_q = 0~101~0~101~0~101~0~101 =21845$$

Subtraindo os valores e aplicando a função $AND$ explicada acima, obtém-se

\begin{equation}
\begin{aligned}
w_{res}& = (w_{node} - w_q) ~AND~ 1~000~1~000~1~000~1~000 \\
& =(48350-21845)~AND~ 1~000~1~000~1~000~1~000 \\
& =26505~AND~ 1~000~1~000~1~000~1~000\\
& = 0~110~0~111~1~000~1~001~~AND~1~000~1~000~1~000~1~000  \\
& = 0\_\_\_0\_\_\_{\bf 1}\_\_\_1\_\_\_  = 136
\end{aligned}
\end{equation}

O primeiro bit 1 é $\lfloor\lg (136)\rfloor=b_7$.
Lembrando que os bits são zero indexados, tem-se 8 bits até o primeiro 1. Dividindo esse valor por 4, que é o tamanho do bloco, obtem-se = 2, que é o penúltimo elemento, ou seja o elemento $\trs{x}=c$ pois $S=(a,b,{\bf c},d)$

\subsection{Ordenando em tempo $o(n \lg n)$}

Já foi visto como realizar uma busca de uma chave com $w$ bits em tempo $O(\frac{\lg n}{\lg \lg n})$ utilizando uma Árvore de Fusão, e como ordenar $n$ elementos utilizando uma árvore B comum. Então, para ordenar $n$ números, basta se inserir elemento por elemento na árvore.~\cite{dynamic} mostra como transformar uma Árvore de Fusão estática, analisada neste artigo, em dinâmica, ou seja, que aceite atualizações nas chaves, fazendo isso em tempo $O(\frac{\lg n}{\lg \lg n} + \lg(\lg(n)))$ por atualização, obtendo assim uma ordenação

\[  \Theta(n\cdot O(\log_w (n) + \lg(\lg(n))) \dfrac{\lg n}{\lg \lg n}) = \]
\[  \Theta(n\cdot \dfrac{\lg n}{\lg \lg n})  \]

\section{Conclusão}
\label{cap:conclusao}

O objetivo deste trabalho foi explicar de maneira detalhada um algoritmo de ordenação que ordena $n$ números em tempo  $O(\frac{n\lg n}{\lg \lg n})$. Para alcançar tal objetivo, verificou-se pouco material disponível sobre o assunto, o que dificultou o  desenvolvimento do trabalho, pois o algoritmo emprega várias definições, teoremas e ideias que não são triviais.

Este trabalho deixa algumas questões complexas em aberto: descobrir o primeiro $b_1$ em uma palavra, calcular o nó $sketch$ em tempo $O(1)$, utilizar Árvores de Fusão dinâmica para inserir os nós na árvore de interesse.

De qualquer forma, este trabalho conseguiu completar com êxito a  analise detalhada  da estrutura de dados da Árvore de Fusão, estrutura fundamental do primeiro algoritmo de ordenação $o(n \lg n)$ e base para muitos algoritmos subsequentes.

Este trabalho também revela que os limites inferiores de tempo precisam ser vistos com muito cuidado. Se um  problema qualquer precisa de, no mínimo, $f(n)$ operações, o verdadeiro limite inferior é $\Omega(f(n)/\lg n)$
pois os modelos computacionais aceitos tem a capacidade de lidar com $\lg n$ de bits em tempo constante.

Um trabalho futuro interessante seria implementar o algoritmo de ordenação baseado em árvores de fusão e comparar 
seu desempenho com algoritmos tradicionais. Outro ponto que merece investigação é a possibilidade de se fazer múltiplas
operações em $O(1)$. Dentro da teoria, poderia-se investigar quais outros problemas poderiam ter sua complexidade 
baixada por meio desta estratégia. Já em computação aplicada, o uso de múltiplas operações dentro de uma palavra
pode acelerar algoritmos tradicionais.

\bibliography{bibl}
\bibliographystyle{plain}


\end{document}